\documentclass[preprint,authoryear,12pt]{elsarticle}

\def\bm#1{\mbox{\boldmath{$#1$}}}

\DeclareMathAlphabet{\bbm}{U}{bbm}{m}{sl}

\newcommand{\code}[1]{\texttt{#1}}

\newcommand\Rey{\mbox{\rm Re}}  
\newcommand\Kn{\mbox{\rm Kn}}  
\newcommand\Mac{\mbox{\rm Ma}} 

\providecommand\bcdot{\boldsymbol{\bm{\cdot}}}
\providecommand\bcol{\boldsymbol{\bm{:}}}

\newsavebox{\astrutbox}
\sbox{\astrutbox}{\rule[-5pt]{0pt}{20pt}}

\usepackage[a4paper,margin=1in,footskip=0.25in]{geometry}
\usepackage{graphicx}
\usepackage{natbib}
\usepackage{subfigure}
\usepackage{booktabs}
\usepackage{amssymb}
\usepackage{amsmath}
\usepackage{wrapfig}
\usepackage{float}
\usepackage{upgreek}
\usepackage{color}
\usepackage[framed,numbered,autolinebreaks,useliterate]{mcode}
\usepackage{enumitem}
\usepackage{amssymb}
\usepackage{amsfonts}
\usepackage{mathtools}
\usepackage{tikz,lipsum,lmodern}
\usepackage[most]{tcolorbox}
\usepackage{scalerel}
\usepackage{stackengine,wasysym}
\usepackage{multirow}

\usetikzlibrary{shapes}

\definecolor{MagentaMat}{RGB}{252,40,252}
\definecolor{PartBlue}{RGB}{230 230 255}

\definecolor{lred}{HTML}{fbf0f2}
\definecolor{lgreen}{HTML}{f1f9f3}
\definecolor{lyellow}{HTML}{fbf8ea}

\begin{document}

\begin{frontmatter}

\title{Modeling high-speed gas--particle flows relevant to spacecraft landings: A review and perspectives}

\author[ME,AE]{Jesse Capecelatro\corref{mycorrespondingauthor}}
\cortext[mycorrespondingauthor]{Corresponding author}
\ead{jcaps@umich.edu}

\address[ME]{Department of Mechanical Engineering, University of Michigan, Ann Arbor, MI 48109-2125, USA}
\address[AE]{Department of Aerospace Engineering, University of Michigan, Ann Arbor, MI 48109-2125, USA}

\begin{abstract}
The interactions between rocket exhaust plumes and the surface of extraterrestrial bodies during spacecraft landings involve complex multiphase flow dynamics that pose significant risk to space exploration missions.
The two-phase flow is characterized by high Reynolds and Mach number conditions with particle concentrations ranging from dilute to close-packing. Low atmospheric pressure and gravity typically encountered in landing environments combined with reduced optical access by the granular material pose significant challenges for experimental investigations.
Consequently, numerical modeling is expected to play an increasingly important role for future missions. 
This article presents a review and perspectives on modeling high-speed disperse two-phase flows relevant to plume-surface interactions (PSI). 
We present an overview of existing drag laws, with origins from 18th-century cannon fire experiments and new insights from particle-resolved numerical simulations. 
While the focus here is on multiphase flows relevant to PSI, much of the same physics are shared by other compressible gas--particle flows, such as coal-dust explosions, volcanic eruptions, and detonation of solid material.
\end{abstract}


\end{frontmatter}


\section{Introduction}
\label{sec:intro}
\begin{figure}[h]
\centering
\includegraphics[width=1.0\textwidth]{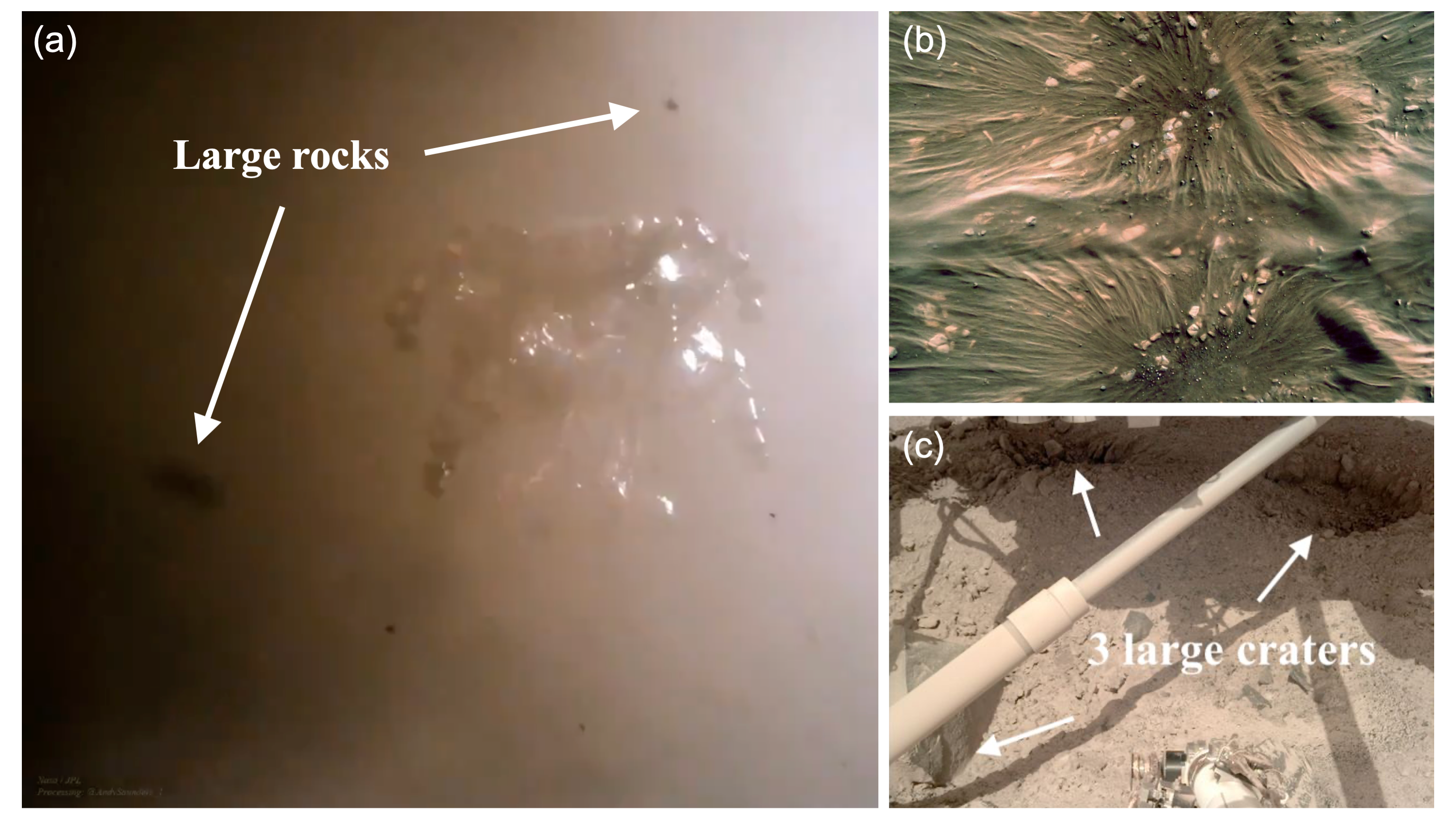}
\caption{Plume-surface interactions during spacecraft propulsive landings represent a key challenge for space exploration. (a-b) Final descent of NASA's Perseverance rover on February 18, 2021 showing (a) ejected dust and rocks and (b) erosion of martian regolith during the Sky Crane maneuver. (c) Craters formed by rocket plumes beneath NASA's InSight lander. Images adapted from NASA/JPL-Caltech.}
\label{fig:PSI}
\end{figure}

During the powered landing of a spacecraft on lunar and planetary bodies, impinging exhaust gases generate a strong recirculation region that fluidize the surface and eject loose granular matter (see Fig.~\ref{fig:PSI}). Plume-surface interactions (PSI) are capable of destabilizing the lander and dislodging dust and debris at high speeds that can damage exposed hardware, reduce visibility, and spoof landing sensors. Mitigating the risks of PSI requires a detailed understanding of the multiphase dynamics under such extreme conditions. While the last several decades have seen significant advancements in simulation and modeling techniques for incompressible particle-laden flows~\citep{crowe1996numerical,balachandar2010turbulent,fox2012large,tenneti2014particle}, much less attention has been paid to high-speed (compressible) two-phase flows. This article presents a review and perspectives on this topic.

In this section, examples of PSI from past landing events and the associated flow physics are presented. The remainder of the article focuses on the fundamental dynamics and processes of compressible gas--particle flows. Origins of existing models and new insights gleaned from particle-resolved simulations are reported. Particular attention is paid to interphase coupling under dilute and moderately dense concentrations at finite Mach number. We summarize classical drag laws in addition to models for unsteady contributions due to added mass and `pseudo' turbulence during shock-particle interactions.

\subsection{Impact of PSI from previous landings}
The detrimental effects of PSI from previous lunar and martian missions are well documented~\citep{christensen1967surveyor,foreman1967interaction,o1970degradation,clark1970effect,jaffe1971blowing,taylor1972apollo,hutton1980surface,o2009direct,gomez2014curiosity}. In fact, four of the six Apollo landings suffered from hindered visibility caused by ejected granular material during landing. Suspension of lunar dust was found to deposit on equipment located 17 m from Apollo 11 and 160 m from Apollo 12~\citep{o2009direct}. Hardware on the Surveyor III craft experienced pitting and cracking as a result of sandblasting from high-speed ejecta during the Apollo 12 landing~\citep{jaffe1971blowing,immer2011apolloa,immer2011apollob}. Particles were estimated to have traveled at speeds in excess of 100 m/s. During the Apollo 15 landing, the lunar module experienced a 12-degree tilt at touchdown, almost terminating the mission~\citep{mcdivitt1971apollo}. Due to the lack of an atmosphere on the Moon, smaller particles suspended during the Apollo landings are estimated to have reached escape velocity, posing hazards to orbital hardware~\citep{lane2008lagrangian}. 

In an effort to mitigate the negative effects of PSI, NASA's 2012 Mars Science Laboratory (MSL) mission introduced the Sky Crane maneuver for the final descent of the Curiosity lander. The rover was lowered down on a bridle from an altitude of 7.5 m to prevent close contact between the rocket engines and the planet's surface. Compared to the Moon, the thin (but finite) martian atmosphere inhibits the spreading of exhaust gas, resulting in collimated rocket plumes that generate highly localized impingement pressures~\citep{mehta2013thruster} and the formation of craters (see Fig.~\ref{fig:PSI}(c)). A significant amount of soil and debris was lifted during MSL, which is believed to have damaged one of the Curiosity rover’s two wind sensors~\citep{gomez2014curiosity}. The same Sky Crane maneuver was also used for the Mars 2020 mission. As shown in Fig.~\ref{fig:PSI}, erosion of martian soil was observed during descent of the Perseverance rover, resulting in the liberation and ejection of large rocks. Meanwhile, planned sample return missions and anticipated crewed missions with higher payloads precludes the use of Sky Crane.

\subsection{Multiphase flow dynamics during PSI}
\begin{figure}[h]
\centering
\includegraphics[width=0.95\textwidth]{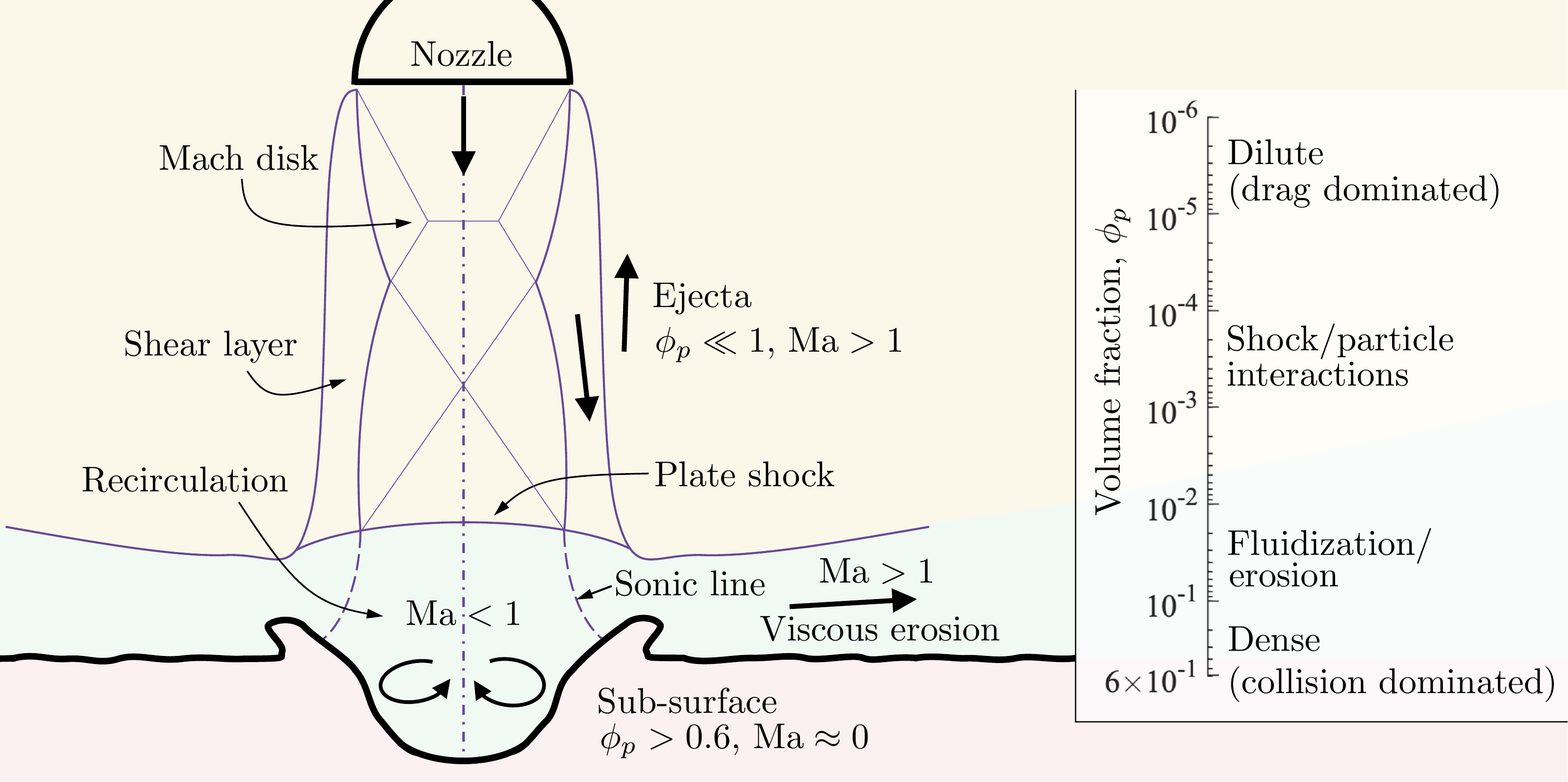}
\caption{Schematic of the fluid-particle dynamics present during a landing event highlighting regions of varying Mach number, $\Mac$, and volume fraction, $\phi_p$.}
\label{fig:PSI2}
\end{figure}

Incorporating the effects of PSI into the design stage of next-generation landers is a necessary step to minimize risk and ensure the success of future missions. What makes this so challenging are the wide range of flow regimes that coexist (see Fig.~\ref{fig:PSI2}). The exhaust gas leaving the rocket nozzle is supersonic and chemically reacting, giving rise to a Mach disk followed by expansion waves and a plate (standoff) shock just above the surface. The Reynolds number at the exit of the nozzle is typically $\Rey=\mathcal{O}(10^5)$ and the Knudsen number of the exhaust gas is sufficiently low~\citep{mehta2013thruster}, thus the flow is continuum and highly turbulent when the lander is within a few meters of the surface.

The particle volume fraction, $\phi_p$, generally decreases with elevation, from near close packing ($\phi_p>0.6$) at the surface to highly dilute above the plate shock. Within the recirculation region (sometimes referred to as the stagnation bubble~\citep{mehta2013thruster}), high-pressure subsonic flow fluidizes the soil. The impinging gas forms shock waves parallel to the surface and deflects the plume radially outward. The deflected plume accelerates to supersonic speeds, inducing high shear stress on the surface that lifts the soil into the boundary layer--so-called viscous erosion~\citep{metzger2009jet}. Ejected particles interact with turbulence in the shear layer and shock waves throughout the exhaust plume where the concentration is low ($\phi_p\ll1$) and the gas-phase Mach number, $\Mac$, is high. Interphase coupling between particles and turbulence in high-speed shear layers are capable of altering pressure fluctuations radiating outwards from the plume~\citep{krothapalli2003turbulence,buchta2019sound} and generating flow instabilities analogous to two-fluid flows \citep{mcfarland2016computational}.



The physical mechanisms contributing to erosion vary depending on the rocket thrust, atmospheric conditions of the landing environment, and physical soil properties. Table~\ref{table:params} lists atmospheric conditions at the surface of Earth, the Moon, and Mars. Regolith (the upper layer of soil) on the Moon is tightly packed due to its near-vacuum conditions, preventing the exhaust gas from penetrating deep within. Thus, erosion is primarily a consequence of shear induced by the viscous flow above the surface~\citep{hutton1968comparison,metzger2009jet}, causing particles to spread far from the lander. Due to the lack of an atmosphere, the Moon is continually bombarded with small meteorites and its regolith exhibits a wide size distribution. In contrast, regolith on Mars is exposed to atmospheric wind and tends to be more spherical~\citep{sullivan2005aeolian}. The stagnation pressure exerted by the jet exceeds the bearing capacity of martian soil and compresses/evacuates the upper layer to form a deep crater~\citep{metzger2009jet}, which redirects the two-phase flow up towards the lander. 

\begin{table}[ht!]
\centering
\caption{Typical atmospheric conditions at the surface of Earth, the Moon, and Mars.}

\begin{tabular}{llll}
\toprule
 & Earth & Moon & Mars \\ 
\midrule
Gravitational acceleration [$\rm{m/s^2}$] &$9.81$ & $1.62$ & $3.72$  \\
Pressure [$\rm{mbar}$] &  $1013.25$ & $3\times10^{-12}$ & $5$--$10$ \\
Density [$\rm{kg/m^3}$] & $1.2$ & $\approx 0$ & $1.66 \times 10^{-2}$ \\
Dynamic viscosity [$\rm{Ns/m^2}$] & $1.8 \times 10^{-5}$ & N/A & $1.5 \times 10^{-5}$  \\
Sound speed [$\rm{m/s}$] &  $343$ & N/A &  $264$ \\
\bottomrule
\end{tabular}
\label{table:params}
\end{table}

\subsection{Modeling PSI}
Theoretical and experimental studies of PSI date back to the early 1960s~\citep{spady1962exploratory,roberts1963action,land1965experimental,roberts1966interface,hutton1969mars}. \citet{roberts1963action} developed the first 
model to describe the erosion and subsequent transport of dust beneath a rocket for lunar environments. The erosion rate was determined from a balance between shear stress exerted by the plume and the threshold shear strength of the soil. 
However, the model neglected aerodynamic forces on the particles, and comparisons to laboratory-scale experiments showed only marginal agreement~\citep{hutton1968comparison}. \citet{metzger2008modification} modified Roberts' theory by incorporating the particle size distribution of lunar soil and imposing ejection angles inferred from Apollo landing videos. However, the authors conclude that a better understanding of the aerodynamic forces and erosion process in supersonic flow regimes are needed. In addition, recent laboratory experiments have revealed that the threshold friction velocity to initiate saltation (and ensuing erosion) differs substantially in low pressure environments like Mars, yet existing scaling laws are unable to capture this~\citep{sullivan2017aeolian,andreotti2021lower}.

Particle motion during PSI is a consequence of fluid-particle and particle-particle interactions at the \textit{microscale}, i.e., fluid stresses and contact dynamics at the sub-particle level. Fluid forces acting on each particle depend non-linearly on the local Reynolds number, Mach number, and volume fraction. Inter-particle forces arising from collisions and sliding/rolling friction vary based on the physical properties of the regolith. In recent years, Eulerian-based two-fluid models~\citep{gale2017gas,balakrishnan2018high,gale2020realistic,balakrishnan2019multi,chinnappan2021modeling,balakrishnan2021fluid} and particle-based methods~\citep{he2012simulation,morris2015approach,rahimi2020near,shallcross2021modeling} of PSI have been conducted that explicitly account for these interactions. Yet, the underlying models these simulations are built upon (e.g., drag, sub-grid scale turbulence, etc.) were largely developed for incompressible flows. While the aerodynamics of high-speed projectiles have been studied for centuries (namely in the context of ballistics), there has been significant developments in recent years, largely owing to the advent of high-performance computing. This article focuses specifically on gas--particle interactions: the origins of existing models; progress in theoretical understanding over recent years; and perspectives for future model development.

\section{Scale separation}
\begin{figure}[h]
\centering
\includegraphics[width=0.5\textwidth]{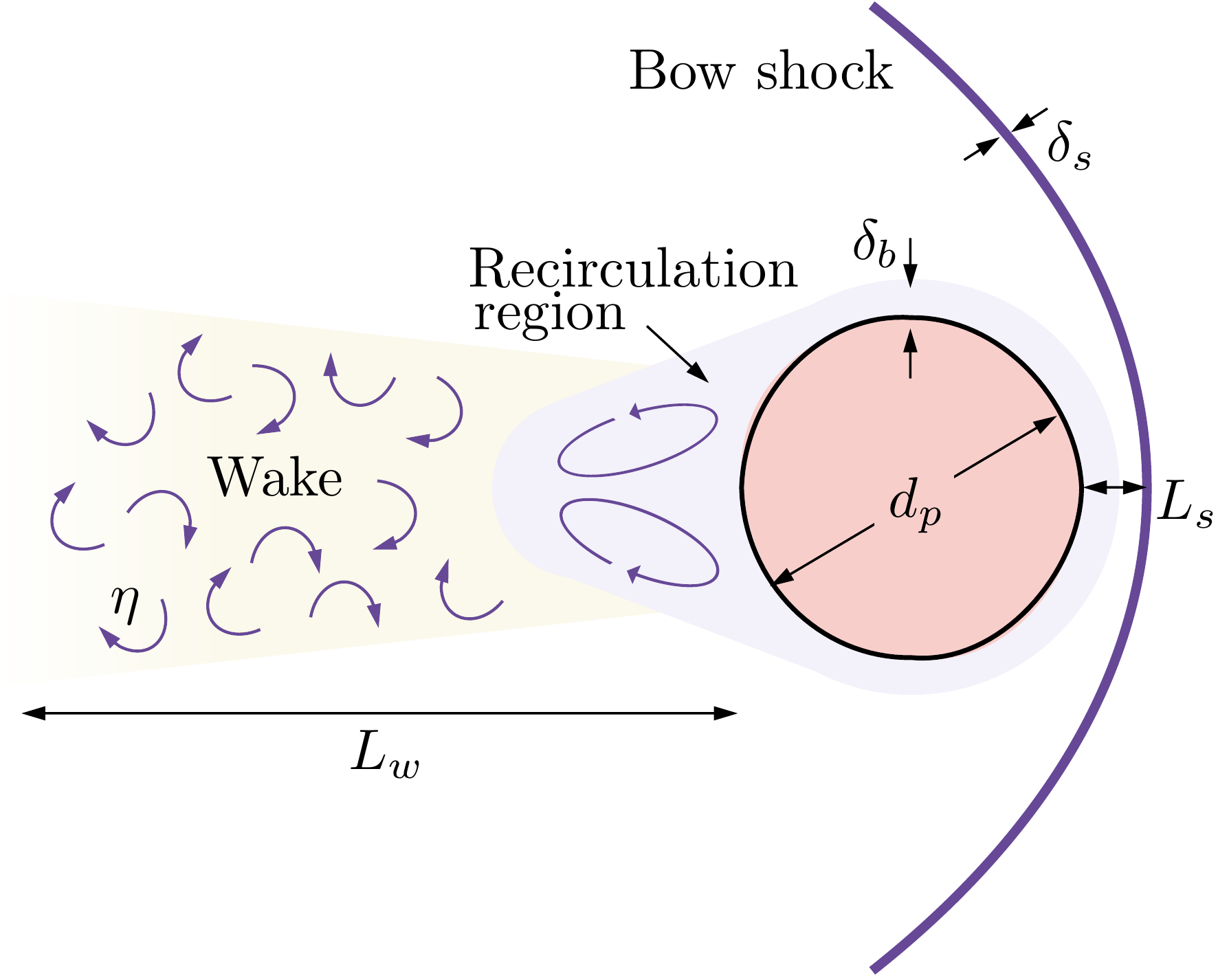}
\caption{Schematic of a supersonic flow past a spherical particle (particle moving from left to right). Various length scales that are present (in approximately increasing magnitude): shock wave thickness $\delta_s$; boundary layer thickness $\delta_b$; Kolmogorov length scale $\eta$; shock standoff distance $L_s$; particle diameter $d_p$; and wake length $L_w$.}
\label{fig:length}
\end{figure}

The range of scales of motion in particle-laden flows are vast (see Fig.~\ref{fig:length} for the simple case of an isolated particle). When the flow is incompressible, the smallest scales are typically on the order of the particle diameter, $d_p$. In turbulent flows, the smallest eddy size (the Kolmogorov, or dissipation length scale) is also important. Relevant time scales include the particle response time, $\tau_p=\rho_p d_p^2/(18\mu_f)$, where $\rho_p$ is the particle density and $\mu_f$ is the fluid viscosity, in addition to time scales associated with the fastest turbulent velocity fluctuations. 
When gas-phase compressibility is important, information travels near or exceeds the sound speed, $c$. An acoustic time scale can be defined using $c$ and a relevant length scale, e.g., $d_p/c$, which is often much smaller than $\tau_p$. For example, a solid particle with density $\rho_p=3000$ kg/m$^3$ and diameter $d_p=100$~$\upmu$m in air results in $\tau_p\approx 0.1$~s and $d_p/c\approx 0.3$~$\upmu$s! The interaction between shock waves and suspensions of particles involves length scales spanning the shock thickness (order of the mean free path of gas molecules), to wakes past particles at scales larger than the particle diameter. In the context of PSI, the largest scales of interest may include the full landing site, which can span kilometers. 

A key challenge in modeling any multiphase flow system is properly capturing the length- and time-scales at play. The equation of motion for a particle with mass $m_p$ and velocity $\bm{v}_p$ traveling through a viscous fluid can be expressed as
\begin{equation}\label{eq:Newton}
    m_p\frac{{\rm d}\bm{v}_p}{{\rm d}t}=\int_S \left(\check{\bm{\sigma}}_f-\check{p}_f\mathbb{I}\right)\bcdot\bm{n}\,{\rm d}S+\bm{F}_{\rm ext},
\end{equation}
where $\check{p}_f$ is the fluid pressure, $\check{\bm{\sigma}}_f$ is the viscous stress tensor, $\bm{n}$ is the unit normal vector outward from the particle surface $S$, $\mathbb{I}$ is the identity tensor, and $\bm{F}_{\rm ext}$ are external body forces. Here, the $\check{(\cdot)}$ notation denotes a microscale quantity prior to any averaging. It becomes immediately apparent that a numerical solution to Eq.~\eqref{eq:Newton} requires sub-particle scale resolution. Particle-resolved simulations for incompressible flows usually employ grid spacing 20--40 times smaller than the particle diameter~\citep{tenneti2014particle}--even finer resolution is typically required for compressible flows--which becomes extremely computationally demanding when the system size of interest is much larger than the size of an individual particle. Thus, direct solutions to the conservation equations are rarely practical. Below we introduce the \textit{averaged} equations of motion and the challenges that arise.

\subsection{Averaged equations}\label{sec:average}
Obtaining a mathematical description that captures the multi-scale nature of two-phase flows typically involves ensemble averaging~\citep{zhang1997momentum} or volume filtering~\citep{anderson1967fluid, capecelatro2013euler}. 
The main idea is to apply an averaging volume (or filter) with a characteristic size comparable to the inter-particle spacing to replace surface fluxes with volumetric source terms. This is analogous to large-eddy simulation (LES) of single-phase flows, and similarly results in unclosed terms that require models. Unlike in single-phase LES, the averaging procedure omits the volume occupied by particles, and consequently differentiation and filtering do not commute. This produces sub-filtered (or subgrid-scale) contributions at the surface of each particle.

Applying such an averaging operation to the viscous compressible Navier--Stokes equations--recall the flow within the exhaust plume during PSI can be treated as continuum--yields equations for mass, momentum, and energy~\citep{shallcross2020volume}
\begin{equation}\label{eq:density}
\frac{\partial \phi_f\rho_f}{\partial t}+\nabla\bcdot\left(\phi_f\rho_f \bm{u}_f\right)=0,
\end{equation}
\begin{equation}\label{eq:momentum}
\frac{\partial \phi_f \rho_f \bm{u}_f}{\partial t} + \nabla \bcdot (\phi_f \left\{\rho \bm{u}_f\bm{u}_f + \bm{R}_f\right\})  =\phi_f\nabla\bcdot \left(\bm{\sigma}_f-p_f\mathbb{I}\right)+\bm{F}_p,
\end{equation}
and
\begin{equation}\label{eq:energy}
\begin{aligned}
\frac{\partial \phi_f \rho_f E_f}{\partial t} & + \nabla \bcdot \left( \phi_f \rho_f E_f \bm{u}_f \right) + \nabla \bcdot (\phi_f (p_f \bm{u}_f - \bm{u}_f \bcdot \bm{\sigma}_f)) + \phi_f \nabla \bcdot \bm{q}_f\\
& = - p_f\frac{\partial\phi_f}{\partial t}+\bm{\sigma}_f\bcol \nabla \left( \phi_p \bm{u}_p \right)  + \bm{u}_p \bcdot \bm{F}_p + Q_p - \nabla \bcdot \left(\phi_f \{ \bm{R}_{p} + \frac{1}{2} \bm{R}_{u} - \bm{R}_{\sigma} \}  \right),
\end{aligned}
\end{equation}
where $\phi_f=1-\phi_p$ is the fluid-phase volume fraction, $\rho_f$ is the fluid density, $\bm{u}_f$ the fluid velocity, $\bm{u}_p$ is the particle-phase velocity (in an Eulerian frame of reference), and $E_f$ the total energy. Interphase coupling takes place through momentum exchange, $\bm{F}_p$, that requires models for drag and other forces acting on the particle (see Sec.~\ref{sec:forces}), and heat exchange, $Q_p$, typically modeled using Nusselt-number correlations. The first term on the right-hand side of Eq.~\eqref{eq:energy} can be thought of as a $pDV$ work term due to particles entering and leaving a control volume~\citep{lhuillier2010multiphase}. By employing the particle-phase continuity equation and assuming constant particle density, this can be replaced with $-p_f\nabla\bcdot\left(\phi_p\bm{u}_p\right)$~\citep{houim2016multiphase}, which may be more convenient in a numerical implementation. 

In Eq.~\eqref{eq:momentum}, $\bm{R}_f$ is a residual stress that arises from filtering the non-linear convective term. Because the fluid velocity fluctuations may originate at the particle scale, the physics that govern this unclosed term differs significantly from the Reynolds stress appearing in classical single-phase flows. In fact, this term may be non-zero even in laminar flows (e.g., via steady wakes), and is therefore termed a \textit{pseudo} turbulent Reynolds stress~\citep{mehrabadi2015pseudo}. While it is typically neglected in incompressible flow models, recent work has shown that $\bm{R}_f$ can contribute to a significant portion of the total kinetic energy during shock-particle interactions \citep{hosseinzadeh2018investigation,sen2018role,mehta2019Pseudo,osnes2019computational,shallcross2020volume}. Special care needs to be taken when distinguishing the sub-filter scale velocity fluctuations originating from large-scale turbulent motion (via a classical energy cascade) and those induced by particles (pseudo turbulence).
This will be revisited in Sec.~\ref{sec:PTKE}.

\begin{figure}[h]
\centering
\includegraphics[width=0.95\textwidth]{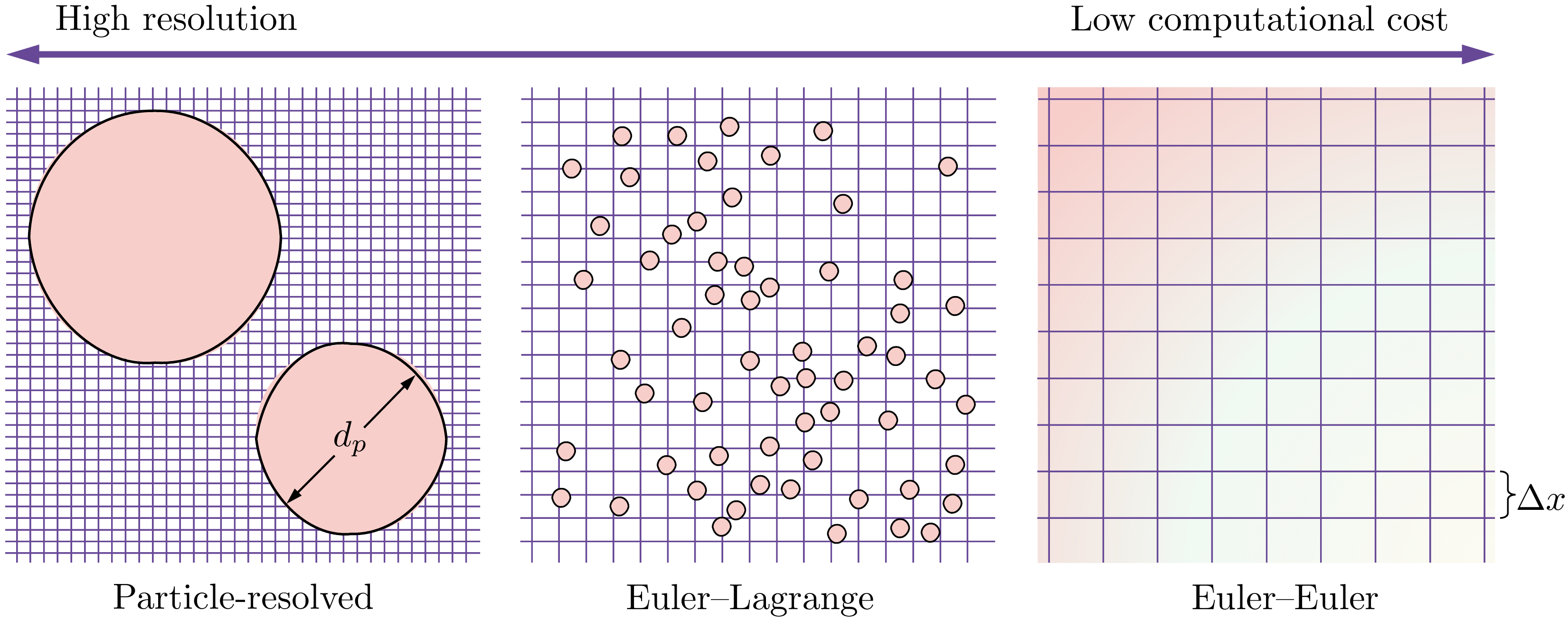}
\caption{Different modeling approached for simulating gas--particle flows. Left: Particle-resolved simulations directly solve the gas-phase equations with grid spacing $\Delta x\ll d_p$. Interphase coupling takes place through boundary conditions at the surface of each particle. At this scale, drag is an \textit{output}. Middle: Euler--Lagrange (EL) methods track individual particles and solve the fluid on a grid with $\Delta x\approx d_p$ (highly-resolved EL) or $\Delta x \gg d_p$ (course EL). Requires closure models for drag and sub-grid scale turbulence. Right: Euler--Euler (EE) methods solve both phases on the grid and require closure models for gas-particle \textit{and} particle-particle interactions.}
\label{fig:multiscale}
\end{figure}

Additional unclosed terms appear in Eq.~\eqref{eq:energy} that account for work due to the subgrid-scale pressure and viscous stress ($\bm{R}_{p}$ and $\bm{R}_{\sigma}$) and a triple product of subgrid-scale velocity fluctuations, $\bm{R}_{u}$. Particle-resolved direct numerical simulations (PR--DNS) of shock waves traveling through arrays of randomly distributed particles have demonstrated that $\bm{R}_{p}$ and $\bm{R}_{\sigma}$ are relatively small compared to the other terms in Eq.~\eqref{eq:energy}~\citep{mehta2019Pseudo,shallcross2020volume}, and \citet{shallcross2020volume} showed that $\bm{R}_{u}\approx2\bm{u}_f\cdot\bm{R}_f$. Thus, accurate solutions of the averaged compressible Navier--Stokes equations namely require validated models for the pseudo-turbulent Reynolds stress, $\bm{R}_f$, and momentum exchange, $\bm{F}_p$. Closure models have been proposed for these terms, but for most part are only valid in the incompressible or dilute limit. This is the focus of Secs.~\ref{sec:forces} and~\ref{sec:PTKE}.

The same averaging procedure could be extended to the particle phase as well, resulting in the so-called Euler--Euler (EE) or two-fluid method. 
This gives rise to even more unclosed terms that require constitutive relations for the solid-phase rheology. While this is outside the scope of the present article, the full set of compressible EE equations can be found in \citet{houim2016multiphase,balakrishnan2018high,fox2019kinetic}. Alternatively, Euler--Lagrange (EL) methods (also referred to as CFD--DEM in the chemical engineering literature) track individual particles and couple them to the averaged fluid equations. Figure~\ref{fig:multiscale} shows an illustration of these different modeling approaches.

\subsection{Ill-posedness and hyperbolicity}\label{sec:ill}
In addition to producing a number of terms that require closure, it is also well established that two-phase models with rigid particles can become ill-posed due to lack of hyperbolicity \citep{lhuillier2013quest}. In general, the compressible EE equations have imaginary characteristics that can lead to unphysical instabilities. This has been shown to give rise to spurious volume fraction disturbances during shock-particle interactions~\citep{theofanous2018shock}. EL methods generally remain hyperbolic due to the Lagrangian treatment of the particle phase, though they could be susceptible to similar numerical instabilities if inter-particle contact is not treated explicitly~\citep{theofanous2017dynamics}. Numerous attempts have been made to restore hyperbolicity starting with \citet{stuhmiller1977influence}, and seems only very recently to be properly resolved~\citep{fox2020hyperbolic}.

Hyperbolicity depends strongly on the nature of the closure models. 
It is well established that the pressure gradient force acting on the particles (termed the Archimedes force) is the source of ill-posedness. Until recently, many authors have resorted to neglecting the Archimedes force or adding an ad-hoc `turbulent dispersion' force to stabilize the solution, which compromises the physical accuracy~\citep{lhuillier2013quest}. Because the Archimedes force is proportional to the density ratio, these issues are more prevalent in bubbly as compared to gas–particle flows. \citet{balakrishnan2021fluid} analyzed the hyperbolicity of a compressible two-fluid model for simulations of supersonic jet-induced cratering on a granular bed. It was found that the system is hyperbolic in most regions except in the vicinity of the crater, particularly where the particle compaction is high ($\phi_p>0.6$). They conclude that high compaction alone is not a sufficient criterion for the system of equations to become degenerate.

An alternative approach to the ad-hoc corrections listed above is to take into account the velocity fluctuations of both phases, which necessitates additional closure for $\bm{R}_f$~ \citep{lhuillier2013quest}. \citet{fox2019kinetic} showed that starting from a well-defined microscale description, it is possible to derive a hyperbolic two-fluid model wherein the fluxes and source terms have unambiguous definitions. Starting from the Boltzmann--Enskog kinetic equations for a binary system, a two-fluid model for fully compressible fluid--particle flows was derived without exclusion of the Archimdes force. A rigorous analysis of this model showed that it is hyperbolic for gas--particle flows with $\rho_p/\rho_f\gg 1$. The following year, \citet{fox2020hyperbolic} extended the model to arbitrary density ratios by including the added mass of the fluid on the particle in addition to fluid-mediated interactions between particles. 
An additional equation to include $\bm{R}_f$ in the fluid phase was also proposed. It was shown that pseudo turbulence has no effect on hyperbolicity but it is needed to ensure conservation. Details on the added mass treatment are given in Sec.~\ref{sec:Fiu} and models for $\bm{R}_f$ are given in Sec.~\ref{sec:PTKE}.

\section{Forces acting on a particle}\label{sec:forces}
As shown in Fig.~\ref{fig:PSI2} and discussed in Sec.~\ref{sec:intro}, PSI spans dilute to dense particle-laden flow regimes. In this section, we first review the state-of-the-art in modeling the forces acting on an \textit{isolated} particle, then move on to extensions to multi-particle systems. In the case of an isolated particle moving through an incompressible fluid at low Reynolds numbers (Stokes flow), \citet{basset1888treatise}, \citet{boussinesq1885application}, and \citet{oseen1927neuere} decomposed the hydrodynamic force (the first term on the right-hand side of Eq.~\eqref{eq:Newton}) into separate contributions: the quasi-steady drag force $\bm{F}_{qs}$; undisturbed flow forces $\bm{F}_{un}$ (sometimes denoted as the pressure gradient or Archimedes force), inviscid unsteady force $\bm{F}_{iu}$ (referred to as added-mass in the limit of incompressible flow), and viscous-unsteady force $\bm{F}_{vu}$ (Basset history). This gives rise to the so-called BBO equation, given by
\begin{equation}\label{eq:BBO}
    m_p\frac{{\rm d}\bm{v}_p}{{\rm d}t}=\underbrace{\bm{F}_{qs}+\bm{F}_{un}+\bm{F}_{iu}+\bm{F}_{vu}}_{\int_S \left(\check{\bm{\sigma}}_f-\check{p}_f\mathbb{I}\right)\bcdot\bm{n}\,{\rm d}S}+\bm{F}_{\rm ext}.
\end{equation}

\citet{parmar2011generalized,parmar2012equation} extended the BBO equations to viscous compressible flows, where the separate force contributions are given by
\begin{subequations}\label{eq:forces}
\begin{align}
\bm{F}_{qs} & = 3\pi\mu_f d_p\left(\bm{u}_f-\bm{v}_p\right)F_D(\Rey_p,\Mac_p,\phi_p),\label{eq:Fqs} \\
\bm{F}_{un} & = V_p\rho_f\frac{{\rm D}\bm{u}_f}{{\rm D}t}, \\
\bm{F}_{iu} & = V_p\int_{-\infty}^t K_{iu}\left(t-\chi;\Mac_p\right)\left(\frac{{\rm D}(\rho_f\bm{u}_f)}{{\rm D}t}-\frac{{\rm d}(\rho_f\bm{v}_p)}{{\rm d}t}\right)_{t=\chi}{\rm d}\chi,\label{eq:Fiu} \\
\bm{F}_{vu} & = \frac{3}{2}d_p^2\sqrt{\pi\rho_f\mu_f}\int_{-\infty}^t K_{vu}\left(t-\chi;\Rey_p,\Mac_p\right)\left(\frac{{\rm D}(\rho_f\bm{u}_f)}{{\rm D}t}-\frac{{\rm d}(\rho_f\bm{v}_p)}{{\rm d}t}\right)_{t=\chi}{\rm d}\chi,\label{eq:Fvu}
\end{align}
\end{subequations}
where $K_{iu}$ is the inviscid unsteady force kernel that will be discussed in detail in Sec.~\ref{sec:Fiu} and $K_{vu}$ is the viscous-unsteady force kernel. The expressions above are valid for an isolated particle in the limit of zero Reynolds number and Mach number. 
Nevertheless, they provide a natural framework for empirical extensions to more realistic flow conditions (though an alternative strategy is proposed in the final section of this article). For example, the quasi-steady drag force includes a correction factor $F_D(\Rey_p,\Mac_p,\phi_p)$ typically given by an empirical correlation based on far-field flow properties. This term is related to the usual definition of the drag coefficient, $C_D$, according to
\begin{equation}
    F_D=\frac{\Rey_p}{24}C_D(\Rey_p,\Mac_p,\phi_p).
\end{equation}
The relevant Reynolds and Mach numbers used in these correlations are based on the relative velocity between the phases, i.e., $\Rey_p=\rho_f\phi_f |\bm{u}_f-\bm{v}_p|d_p/\mu_f$ and $\Mac_p=|\bm{u}_f-\bm{v}_p|/c$, where $\phi_f |\bm{u}_f-\bm{v}_p|$ is the superficial velocity that characterizes the Reynolds number at finite volume fraction. However, it remains an open question how best to interpret the fluid quantities at the particle location in a numerical simulation, especially for flows involving shocks where the flow state can exhibit steep gradients. Simplifying assumptions can also be made to the unsteady force contributions and modeled using empirical correlations. In the following sections we summarize existing models for quasi-steady drag, the inviscid unsteady force, their origins, and extensions to multi-particle systems.

We briefly note that to date, little attention has been paid to finite Mach number and volume fraction corrections for $\bm{F}_{vu}$, and its relative importance in high-speed gas--particle flows is not clear (and is likely small at high Reynolds numbers~\citep{ling2011importance}). Thus, we refrain any further discussion towards this term.

\subsection{Quasi-steady drag on an isolated particle}\label{sec:iso}

The reader already versed in multiphase flow will be familiar with the seminal work by George Stokes who derived an analytic solution to the drag force on a sphere in the limit $\Rey_p\rightarrow0$ in 1851~\citep{stokes1851effect}. The resulting drag coefficient, $C_D=24/\Rey_p$, acts as the basis for nearly all drag laws that have been proposed thereafter. Yet, the contributions by Francis Bashforth the following decade are far less known, despite its influence on modern day compressible drag formulations. 
The astute reader might recognize the name, as he also had a hand in developing the Adams--Bashforth method in 1883~\citep{bashforth1883attempt}, a class of multi-step methods commonly used today for numerical time integration. But it was his invention of the ballistic chronograph in 1864 that provided the first reliable data on the aerodynamics of high-speed projectiles. 

Bashforth's experiments used artillery round shots (cannon fire) that allowed for up to 10 velocity measurements to be made per shot~\citep{bashforth1870}. The size and speed of cannonballs offers a 
surprisingly useful flow regime for studying the drag force. Typical velocities span $100-700$ m/s, corresponding to $0.3\le\Mac_p\le2$ in air, over which $C_D$ changes sharply due to compressibility effects. In addition, the Reynolds numbers straddle the critical value ($\Rey_p\approx2\times10^5$) where $C_D$ exhibits an abrupt drop. Bashforth showed that the drag force at moderate (subsonic) velocities is nearly proportional to the square of velocity, at greater velocities (subsonic to supersonic) it is nearly proportional to the cube of velocity, and at even higher velocities it is again closely proportional to the square of velocity~\citep{gilman1905ballistic}. These experiments were some of the last to use round shots--artillery fire at that time transitioned to elongated cylindrical bullets--and consequently are among few measurements ever made of high-speed, relatively large ($74-225$ mm) spheres.

More than a century later, \citet{miller1979sphere} compiled available data for the drag on
spheres over a wide range of Reynolds and Mach numbers. Interestingly, the
most accurate high Reynolds number data for moderate to high Mach numbers turned out to be from Bashforth's experiments. They showed that his measurements scatter no more than modern data at that time, which was used to construct a comprehensive map of $C_D$ as a function of $\Rey_p$ for $0.1\le\Mac_p\le3$. However, they concluded that significant scatter exists among all the data collected. As stated in \citet{clift2005bubbles}: \textit{``Many of the data in the literature for $\Mac>0.2$ are unreliable \ldots because of high levels of freestream turbulence, interference by a support, wall effects, etc."} 
This is not surprising, as much of this data is based on 18-th and 19-th century cannon firings!

Using available experimental and theoretical data at that time, \citet{henderson1976drag} developed a Reynolds- and Mach-number dependent drag correlation as piecewise functions for subsonic, supersonic ($\Mac_p>1.75$), and linearly interpolated transonic regimes (see \ref{app:henderson}). It includes non-continuum effects when the mean-free path of the gas phase $\lambda$ approaches the particle diameter, i.e., finite values of the Knudsen number $\Kn_p=\lambda/d_p$. For an ideal gas, the Knudsen number can be defined in terms of the Reynolds and Mach number according to \citep{clift2005bubbles}
\begin{equation}
    \Kn_p=\sqrt{\frac{\pi\gamma}{2}}\frac{\Mac_p}{\Rey_p},
\end{equation}
where $\gamma$ is the ratio of specific heats for the gas phase. When $\Kn_p<10^{-3}$ the fluid can be treated as a continuum. At larger values of the Knudsen number, the collision rate of gas molecules with the surface of the particle is not high enough to satisfy the usual no-slip condition. At moderate but still small values, the flow exhibits a small departure from no-slip (slip regime). When $\Kn_p>10$, gas molecules collide with the particle but inter-molecule interactions are rare (free molecular flow). These regimes are summarized in Table~\ref{table:Kn}.

\begin{table}[h]
\centering
\caption{Flow regimes based on the Knudsen number~\citep{schaaf1958flow}.}
\begin{tabular}{lr}
\toprule
Knudsen number range & Flow regime\\
\midrule
$\Kn_p\le0.01$ & Continuum regime\\
$0.01<\Kn_p\le0.1$ & Slip flow\\
$0.1<\Kn_p\le10$ & Transitional flow\\
$\Kn_p>10$ & Free molecular flow\\
\bottomrule
\end{tabular}
\label{table:Kn}
\end{table}

It was later shown by \citet{loth2008compressibility} that the drag coefficient at finite Mach numbers can be separated into two regimes: a rarefaction-dominated regime at low Reynolds numbers ($\Rey_p\lessapprox45$); and a compression-dominated regime at higher Reynolds numbers ($\Rey_p\gtrapprox45$). In between, it was suggested that the competing effects cancel each other out, leading to a so-called nexus condition, where $C_D\approx1.63$ is independent of $\Mac_p$ and $\Kn_p$. Two separate models were constructed for these regimes. The drag model is based on theoretical analysis and experimental data from \citet{hoerner1958fluid} taken from ballistic ranges.  Similar to the model by \citet{henderson1976drag}, it relies on interpolation and approximation due to a lack of reliable data in the transitional regime ($\Mac_p\approx0.9$).

\begin{figure}[h]
\subfigure[$\Mac_p=0.9$]{\includegraphics[width=0.32\textwidth]{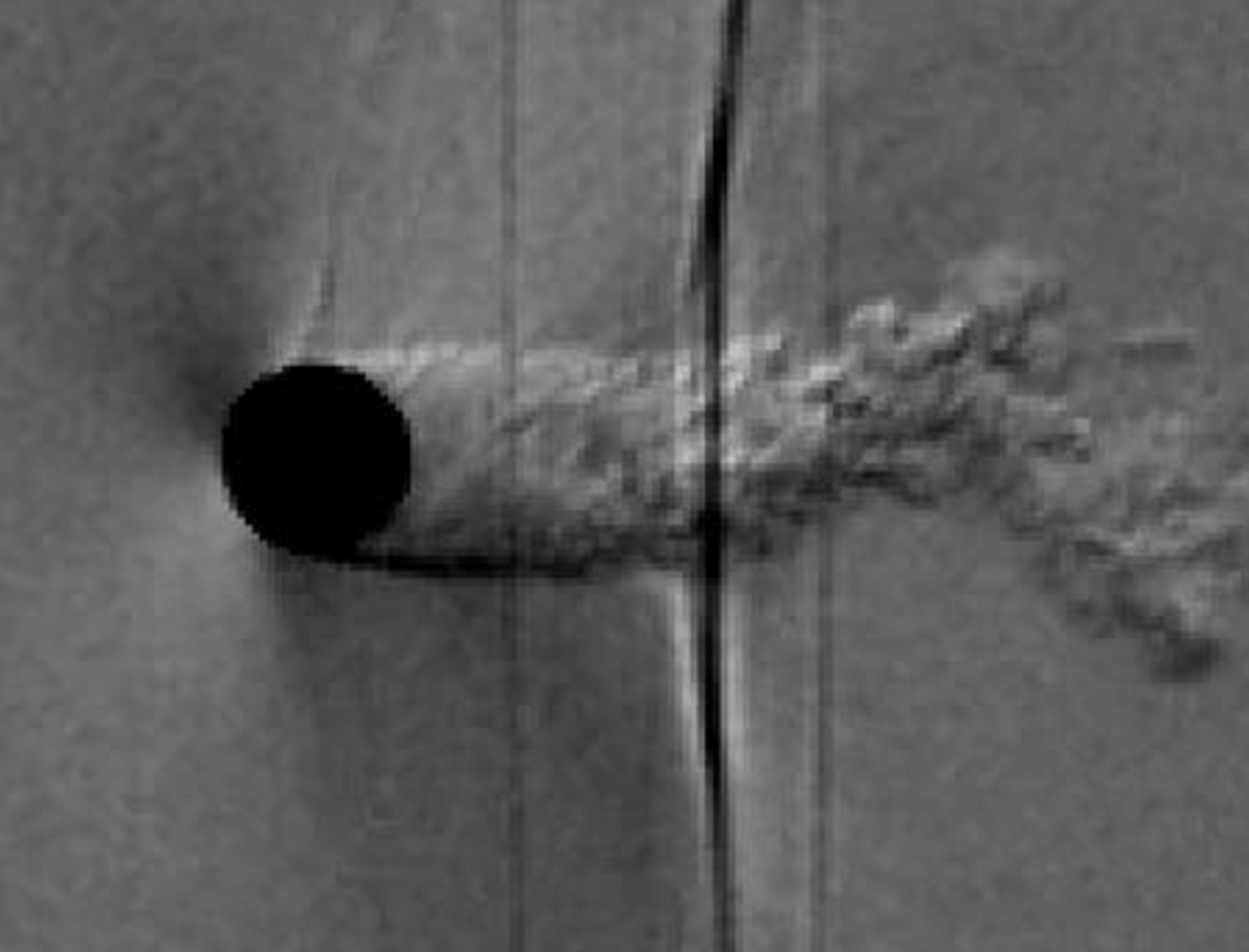}}
\subfigure[$\Mac_p=1.21$]{\includegraphics[width=0.32\textwidth]{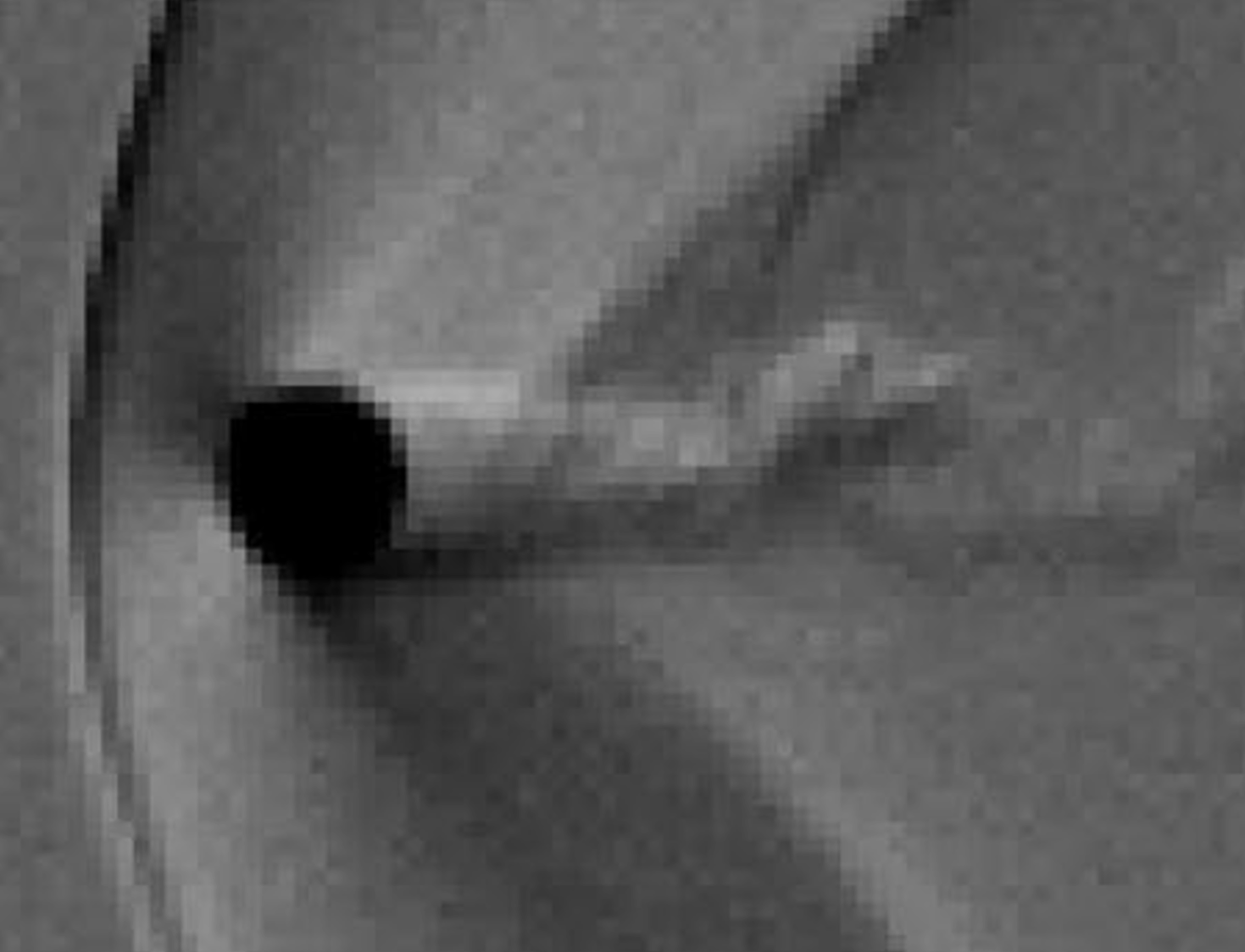}}
\subfigure[$\Mac_p=1.39$]{\includegraphics[width=0.32\textwidth]{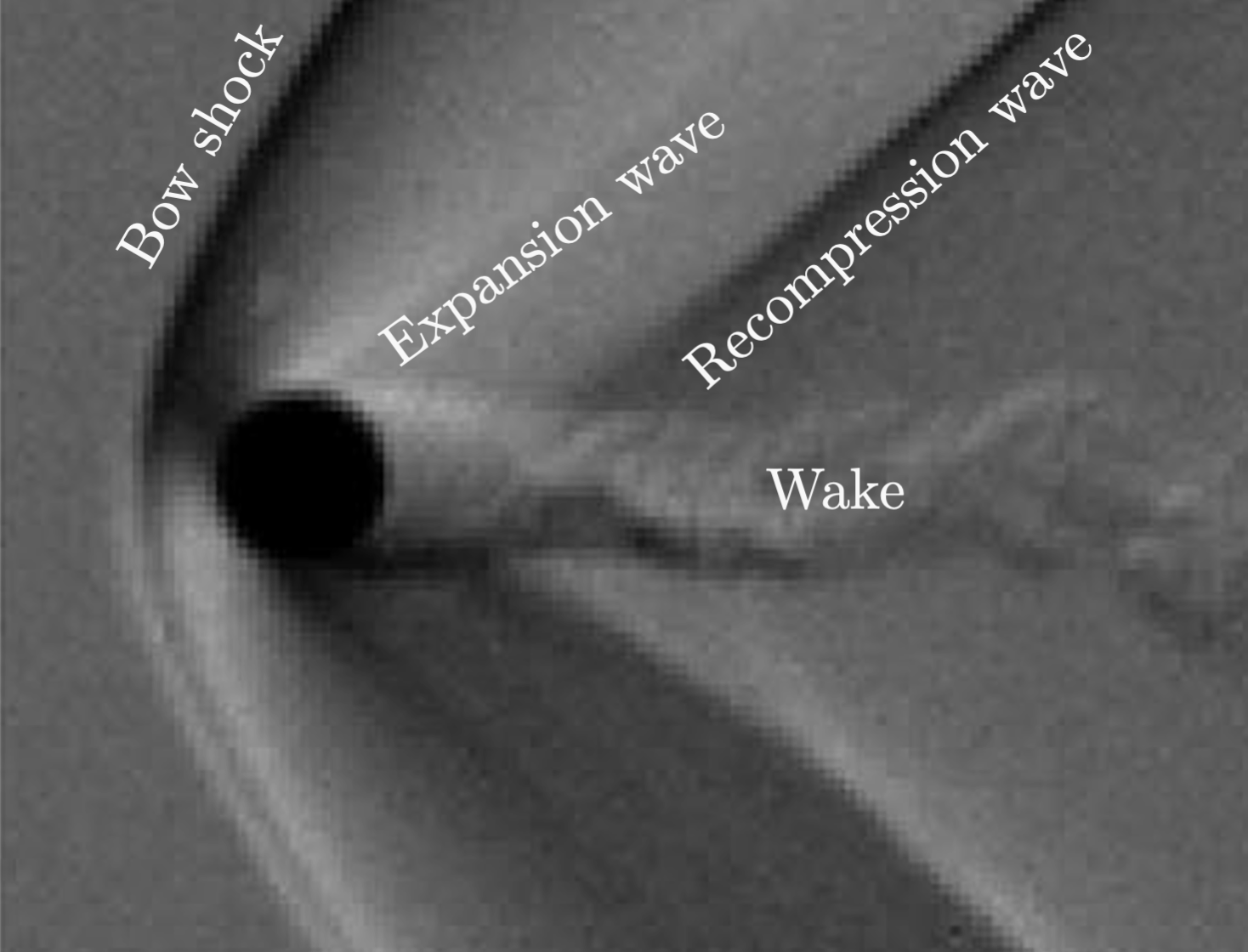}}
\centering
\caption{Schlieren visualization of flow past a sphere with $\Rey_p=\mathcal{O}\left(10^5\right)$. Adapted from \citet{nagata2020experimental} with permission.}
\label{fig:Nagata}
\end{figure}

\citet{parmar2010improved} assessed the models of \citet{henderson1976drag} and \citet{loth2008compressibility} using data collected by \citet{bailey1976sphere}, and proposed an improved correlation for the drag coefficient. The model by \citet{henderson1976drag} was found to exhibit a maximum deviation from the data of 16\%. The correlation by \citet{loth2008compressibility} was found to deviate by as much as 55\% when $\Mac_p\approx 0.9$. The model proposed by \citet{parmar2010improved} was developed for continuum flows ($\Kn<0.01$) with $\Rey_p\le 2\times 10^5$ and ${\rm  Ma}_p\le 1.75$. The drag coefficient is decomposed into three correlations for subcritical ($\Mac_p<\Mac_{\rm cr}\approx0.6$), intermediate, and supersonic Mach number regimes, according to
\begin{equation}\label{eq:Parmar}
C_D(\Rey_p,\Mac_p)=
    \begin{cases}
      C_{D,{\rm std}}(\Rey_p)+\left[C_{D,{\rm cr}}(\Rey_p)-C_{D,{\rm std}}(\Rey_p) \right]\frac{\Mac_p}{\Mac_{\rm cr}} & \text{if } \Mac_p<\Mac_{\rm cr},\\
      C_{D,{\rm sub}}(\Rey_p,\Mac_p) & \text{if } \Mac_{\rm cr}<\Mac_p\le 1,\\
       C_{D,{\rm sup}}(\Rey_p,\Mac_p) & \text{if } 1<\Mac_p\le 1.75.
    \end{cases}
\end{equation}
In the limit of zero Mach number, the drag coefficient reduces to $C_{D,{\rm std}}$, the standard correlation of \citet{clift1970motion}  valid for $\Rey_p<2\times10^5$ (given in \ref{app:CG}). 
The Mach number-dependent correlations are provided in \ref{app:parmar}. For subcritical Mach numbers, the drag coefficient is only weakly affected by compressibility effects due to the absence of any shock (or expansion) waves. For supercritical but still subsonic Mach numbers, $C_D$ becomes more strongly dependent on the Mach number due to the presence of a weak shock. For supersonic Mach numbers, a bow shock is formed (with a standoff distance that decreases with increasing Mach number) that leads to a large increase in $C_D$. Schlieren visualization of a sphere in free flight from recent experiments by \citet{nagata2020experimental} are shown in Fig.~\ref{fig:Nagata}, highlighting these flow regimes.

With the advent of high-performance computing, DNS of compressible particle-laden flows is beginning to shed new light on this topic. \citet{nagata2016investigation,nagata2020direct} performed DNS of flow past an isolated sphere for $0.3\le\Mac_p\le2$ and $250\le\Rey_p\le1000$. They report shortcomings of existing drag correlations owing to the compromised accuracy in the transitional regime ($0.9<\Mac_p<1$) due to lack of experimental data. They demonstrated that the effect of $\Mac_p$ and $\Rey_p$ on the flow structure and drag coefficient can be characterized by the position of the separation point. A rapid extension of the length of the recirculation region was observed in the transitional regime, where $C_D$ was shown to increase with increasing Mach number independently of the Reynolds number.

\begin{figure}[h]
\centering
\includegraphics[width=0.72\textwidth]{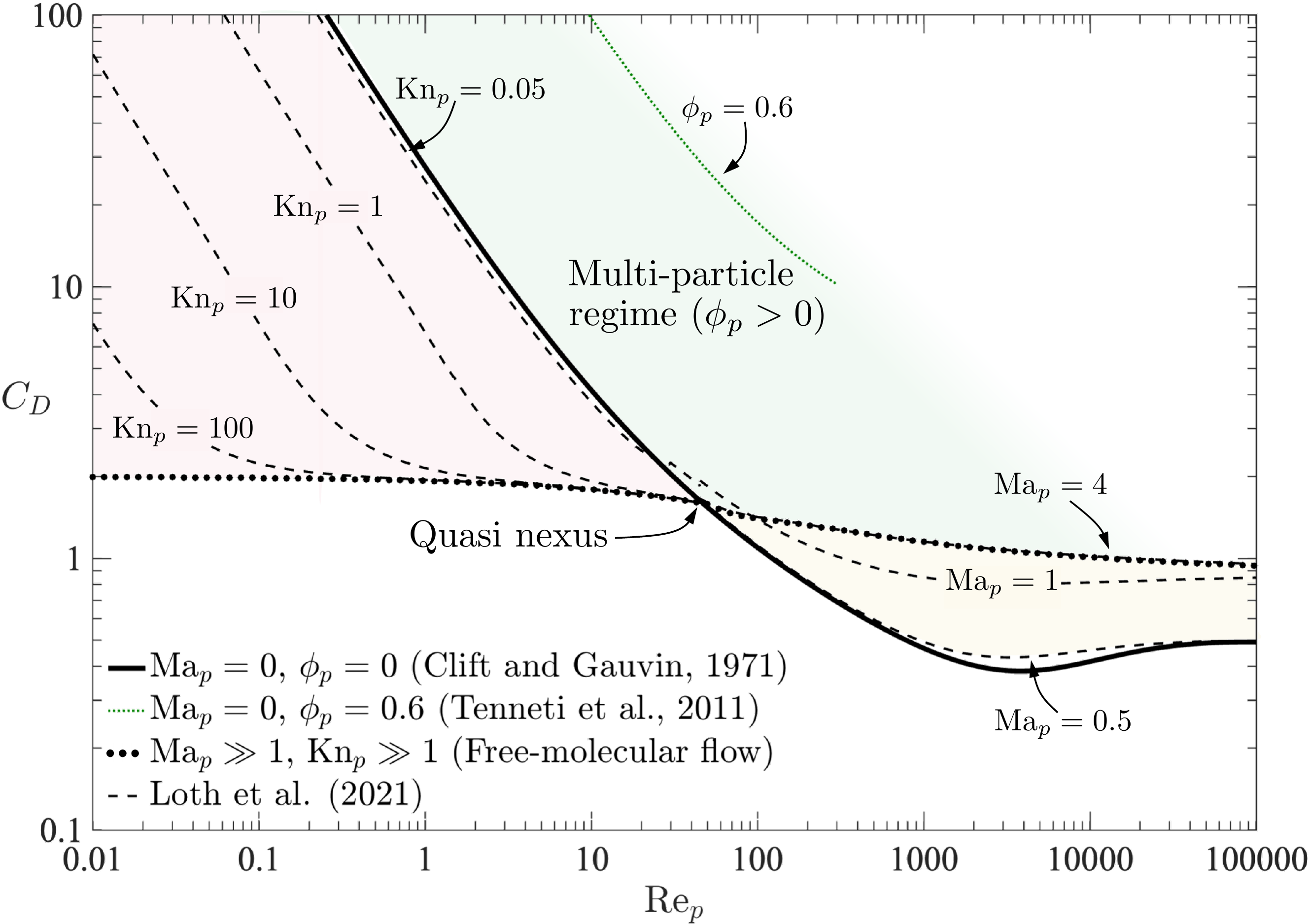}
\caption{Drag coefficient on a spherical particle showing rarefaction (shaded in red) and compression (shaded in yellow) dominated regimes using the new model proposed by \citet{loth2021supersonic}. Multi-particle effects on the drag coefficient are shown in green.}
\label{fig:drag}
\end{figure}

Shortly after, \citet{loth2021supersonic} combined the DNS data of \citet{nagata2016investigation,nagata2020direct} with rarefied-gas simulations and an expanded experimental dataset to develop new empirical models for the drag coefficient (given in \ref{app:loth}). It is simpler than that originally proposed by \citet{loth2008compressibility} and appears to be the most accurate and comprehensive model developed for compressible gas--particle flows to date (and the first to incorporate numerical data). It was shown to be approximately twice as accurate compared to the correlations of \citet{loth2008compressibility} and \citet{parmar2010improved} at moderate Mach numbers, and showed improvement to Loth's original model at higher Mach numbers ($\Mac_p>2$) and for rarefied conditions. The overall trends in $C_D$ are shown in Fig.~\ref{fig:drag}, highlighting the quasi-nexus that separates the rarefaction and compressibility flow regimes. It can be seen that the compression-dominated region ($\Rey_p>60$) yields an increase in $C_D$ as $\Mac_p$ increases, whereas in the rarefaction-dominated region ($\Rey_p<30$), $C_D$ is inversely proportional to $\Mac_p$. In between these regimes--the quasi-nexus--there exists a weak transonic bump that connects the two. While DNS was used to quantitatively describe this behavior, the authors suggest additional data is warranted to further refine the drag coefficient within the quasi-nexus region.

\subsection{Drag in multi-particle systems}
\begin{figure}[h]
\subfigure[$\Rey_p=20$, $\phi_p=0.1$, $\Mac_p=0$]{\includegraphics[width=0.42\textwidth]{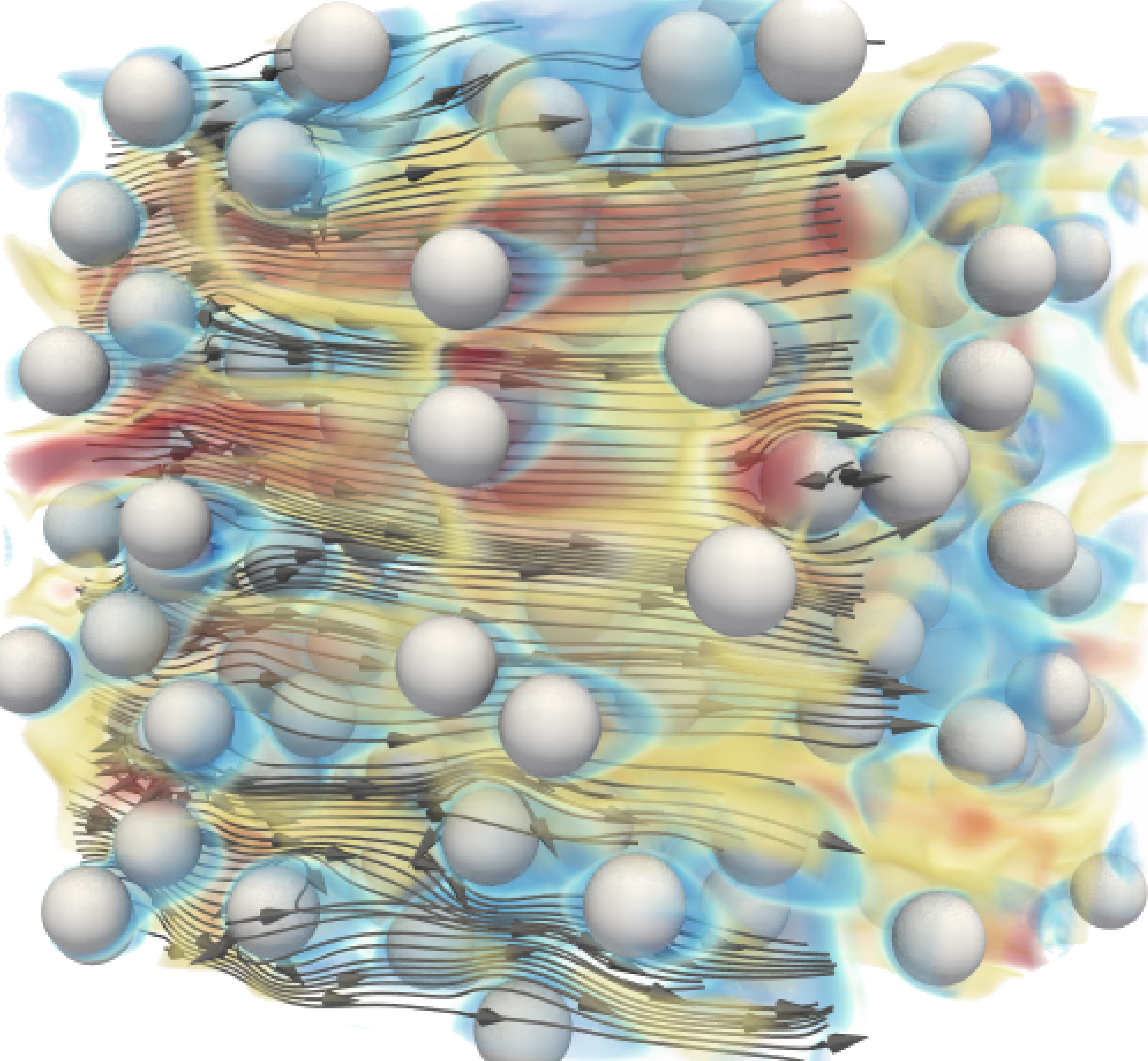}}\quad\quad
\subfigure[$\Rey_p=300$, $\phi_p=0.1$, $\Mac_p=0.8$]{\includegraphics[width=0.39\textwidth]{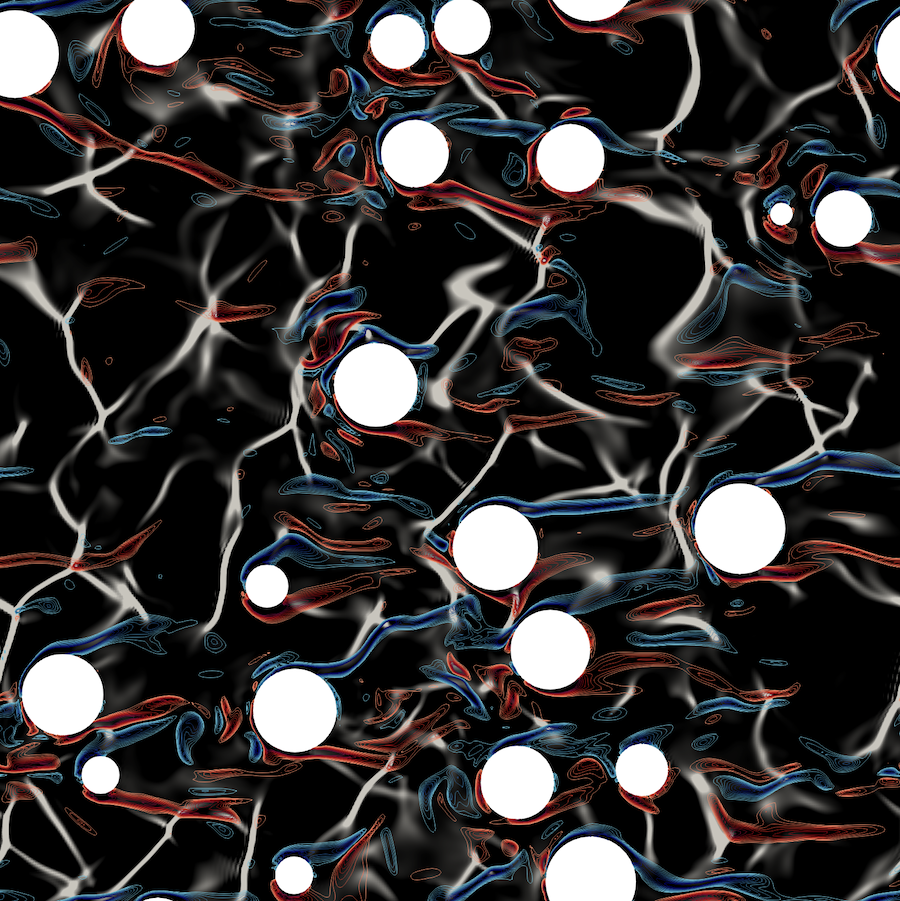}}
\centering
\caption{Instantaneous snapshots from PR--DNS of a homogeneous suspension of particles. (a) Incompressible flow with $\rho_p/\rho_f=1000$ showing fluid velocity (red: high, blue: low)~\citep{lattanzi2021fluid}. (b) Compressible flow past fixed spheres showing a 2D cross-section of the 3D flow highlighting fluid dilatation (black/white) and vorticity (red: positive, blue: negative)~\citep{Khalloufi2021}.}
\label{fig:PRDNS}
\end{figure}

In the presence of one or more neighbors, the drag force exerted on a particle can vary significantly compared to that of an isolated sphere. Multi-particle drag correlations are needed when the inter-particle distance becomes comparable to the particle diameter--typically when $\phi_p\gtrapprox 1\times10^{-3}$. In a homogeneous fluidization system (see Fig.~\ref{fig:PRDNS}), the single-particle drag coefficient can be corrected to take into account finite volume fraction effects. In general, the mean drag over a suspension of particles has a complex dependency on $\Rey_p$ and $\phi_p$ (and likely on $\Mac_p$ as well, though this is far less studied).

Correlations used to account for the drag force in multi-particle systems can be classified into two types. In the first type, \citet{ergun1952fluid} proposed a correlation based on the drag force in the limit of Stokes flow with an additional term that depends linearly on $\Rey_p$, such that the drag correction in Eq.~\eqref{eq:Fqs} is expressed as
 \begin{equation}\label{eq:Ergun}
     F_D(\Rey_p,\phi_p)=F_D(0,\phi_p)+\alpha\Rey_p.
 \end{equation}
Originally, $\alpha$ was only a function of volume fraction~\citep{ergun1952fluid}. However, it was later shown that $\alpha$ also depends non-linearly on $\Rey_p$~\citep{hill2001first,beetstra2007drag,tenneti2011drag,tang2015new}. Recently, \citet{tang2016direct} proposed a drag law of this type for freely-evolving spherical particles by incorporating the granular temperature (a measure of the particle velocity variance) to account for particle mobility effects (see \ref{app:tang}). They found that particle mobility increases the drag force compared to stationary particles at the same volume fraction, with greater effect at higher $\Rey_p$.
 
The second correlation type originally proposed by \citet{wen1966mechanics} takes the form
\begin{equation}\label{eq:WenYu}
    F_D(\Rey_p,\phi_p)=F_D(\Rey_p,0)\phi_f^{-\beta},
\end{equation}
where $F_D(\Rey_p,0)$ is the finite Reynolds number drag correction for an isolated particle (see Sec.~\ref{sec:iso}). The value for $\beta$ was originally constant \citep{wen1966mechanics}, but was later shown to depend on $\Rey_p$ \citep{di1994voidage}. More recently, \citet{tenneti2011drag} proposed a model following this form based on PR--DNS of flow past random assemblies of fixed spherical particles valid for $0.01\le\Rey_p\le 300$ and $0.1\le\phi_p\le0.5$ (given in \ref{app:tenneti}). This was later extended to freely evolving particles for $0.001\le\rho_p/\rho_f\le1000$~\citep{tavanashad2021particle}, where it was shown that $\beta$ depends on the density ratio. Contrary to what \citet{tang2016direct} observed, this data showed that the drag force for freely-evolving particles approaches that of fixed particle assemblies when $\rho_p/\rho_f>100$. Recall $\rho_p/\rho_f=\mathcal{O}\left(10^5\right)$ during PSI.

Compared to incompressible flows, much less attention has been paid to multi-particle drag correlations for flows at finite Mach number. The presence of neighboring particles acts to increase the drag coefficient compared that of an isolated particle. At moderate Reynolds numbers, this increase in drag coefficient obfuscates the quasi-nexus region shown in Fig.~\ref{fig:drag}. A third, multi-particle dominated regime exists in the upper-right quadrant of Fig.~\ref{fig:drag}, located above the \citet{clift1971motion} curve. This may intersect with the compression-dominated regime, though its precise behavior remains largely unknown.

In an attempt to incorporate volume fraction effects into compressible drag correlations, \citet{ling2012interaction} combined the Mach number-dependent drag model of \citet{parmar2010improved} with the (incompressible) volume fraction correction of \citet{sangani1991added}. Note this follows the second correlation type given by Eq.~\eqref{eq:WenYu}.
The corresponding drag coefficient is given by
\begin{equation}
C_D(\Rey_p,\Mac_p,\phi_p)=C_{D,{\rm std}}(\Rey_p)\xi_1(\Rey_p,\Mac_p)\xi_2(\phi_p),
\end{equation}
where $\xi_1(\Rey_p,\Mac_p)$ is the Mach number correlation for an isolated particle that was provided in Eq.~\eqref{eq:Parmar}, and the volume fraction correction is given by
\begin{equation}
    \xi_2(\phi_p)=\frac{1+2\phi_p}{\left(1-\phi_p\right)^2}.
\end{equation}
The above expression was analytically derived from periodic arrays of spherical particles in the zero Reynolds number limit for $\phi_p\le 0.3$ \citep{sangani1991added}. 
However, such an approach does not guarantee it will capture the non-linear coupling between $\Rey_p$, $\Mac_p$, and $\phi_p$ in general. Recent PR--DNS of shock-particle interactions have shown that particles experience a large variation in drag force due to fluid-mediated particle-particle interactions, which current models fail to capture
\citep{mehta2018propagation,mehta2019effect,osnes2021performance}.

While PR--DNS of multi-particle systems are now common for incompressible flow, only in recent years PR--DNS of compressible flows past assemblies of particles have come online \citep[e.g.,][]{regele2014unsteady,mehta2016shock,mehta2018propagation,theofanous2018shock,hosseinzadeh2018investigation,osnes2019computational,shallcross2020volume}. However, these studies typically consider a shock wave traveling through an array of particles, which introduces challenges in developing drag correlations due to the lack of statistical stationarity and homogeneity. Figure~\ref{fig:PRDNS}(b) shows an example of PR--DNS of a homogeneous flow with $\phi_p=0.1$ and $\Mac_p=0.8$ based on the mean slip velocity \citep{Khalloufi2021}. The flow accelerates to supersonic speeds in the interstitial space between particles, resulting in large values of fluid dilatation (shocklets). These regions of high compression have important consequences on vorticity generation due to baroclinic torque ($\nabla\rho_f\times\nabla p_f$) at the vertex of curved shocks \citep{kida1990enstrophy}, which is anticipated to alter the drag force. This work among other recent PR--DNS represent a promising new direction for developing improved drag correlations that span relevant values of $\Rey_p$, $\Mac_p$, and $\phi_p$ for applications related to PSI and other compressible gas--particle flows.

\subsection{Unsteady inviscid force}\label{sec:Fiu}
Compared to the quasi-steady drag force, the unsteady forces appearing in Eq.~\eqref{eq:forces} have received much less attention, especially in the context of compressible flows. It is typically assumed that these terms scale with the fluid-to-particle density ratio, and therefore can be neglected for gas--solid flows. This is generally true for the case of a particle accelerating in a quiescent fluid. However, when the surrounding fluid is accelerating, the order of magnitude of the inviscid unsteady terms ($\bm{F}_{iu}$ and $\bm{F}_{un}$) relative to the quasi-steady drag force scales like $\Rey_p d_p/L_f$, where $L_f$ is the characteristic length scale of the background flow \citep{bagchi2002steady}. Therefore, the motion of finite-size particles in a compressible flow undergoing strong acceleration (e.g., in the presence of a shock wave) can be greatly influenced by the unsteady forces, regardless of the density ratio. In fact, the peak drag coefficient of an isolated particle interacting with a shock wave can be more than 10 times larger than the value from the quasi-steady drag force depending on the shock Mach number, $\Mac_s$, and $\Rey_p$ \citep{parmar2009modeling,ling2011importance}.

The inviscid unsteady force given in Eq.~\eqref{eq:Fiu} is expressed in terms of a response kernel, $K_{iu}(\tau;\Mac_p)$, used to weigh the history of the particle's acceleration. The kernel decays over a non-dimensional acoustic time $ct/d_p$, which depends on both the particle's shape and Mach number. In an incompressible flow, the acoustics propagate infinitely fast and the kernel reduces to a Dirac delta function, i.e., $K_{iu}(\tau;\Mac_p=0)=\delta(\tau)/2$. 
For a spherical particle in an incompressible flow, integration of the kernel results in an added mass coefficient $C_m=0.5$. However, when the flow is compressible, the Mach number delays the approach to steady state and the force no longer takes the form of a constant mass multiplied by the instantaneous acceleration. Because of this, \citet{miles1951virtual,longhorn1952unsteady} and later reiterated by \citet{parmar2008unsteady}, emphasized that reference to this force as a `virtual' or `added' mass is only applicable to incompressible flows.

A popular expression for $K_{iu}$ was obtained by \citet{longhorn1952unsteady} for a spherical particle in the limit of zero Mach number as $K_{iu}(\tau;\Mac_p=0)=\exp(-\tau)\cos(\tau)$. More than half a century later, \citet{parmar2008unsteady} extended this to finite (but subcritical) Mach numbers. They demonstrated that $\Mac_p$ has a pronounced effect on both the peak value of the unsteady force and the effective added-mass coefficient. By employing the compressible form of the Bernoulli equation, a Mach number expansion for pressure was derived and integrated around a sphere to obtain the force. 
At a Mach number of $0.5$, the effective added-mass coefficient was found to be $C_m(\Mac_p=0.5)\approx 1$, approximately twice as large as the incompressible value for a sphere.

In general, $K_{iu}$ becomes negligibly small after a few acoustic time scales. If the relative acceleration changes slowly over this duration, it can be taken outside of the integral and Eq.~\eqref{eq:Fiu} can be rewritten as
\begin{equation}\label{eq:Fiu2}
    \bm{F}_{iu}  \approx V_p C_m(\Mac_p,\phi_p)\left(\frac{{\rm D}(\rho_f\bm{u}_f)}{{\rm D}t}-\frac{{\rm d}(\rho_f\bm{v}_p)}{{\rm d}t}\right),
\end{equation}
where $V_p$ is the particle volume. \citet{ling2012interaction} proposed a model for the added mass coefficient, given by
\begin{equation}
C_m(\Mac_p,\phi_p)=C_{M,{\rm std}}\eta_1(\Mac_p)\eta_2(\phi_p),
\end{equation}
which reduces to the usual value $C_{M,{\rm std}}=0.5$ in the limit of zero Mach number and volume fraction. The Mach number correction (valid in the subcritial regime) is given by~\citep{parmar2008unsteady}
\begin{equation}
\eta_1(\Mac_p)=
    \begin{cases}
      1+1.8\Mac_p+7.6\Mac_p^4 & \text{if } \Mac_p<0.6,\\
      2.633 & \text{otherwise}.
    \end{cases}
\end{equation}
The volume fraction correction is taken from the expression by \citet{sangani1991added} as
\begin{equation}
    \eta_2(\phi_p)=\frac{1+2\phi_p}{1-\phi_p},
\end{equation}
which is valid for random arrays of spheres in incompressible low Reynolds number flow when $\phi_p<0.3$. Such a simple expression for the inviscid unsteady force has received mixed success when applied to shock-particle interactions \citep{parmar2009modeling,koneru2021assessment,osnes2021performance}.

An interesting alternative to modeling the inviscid unsteady force was recently proposed by \citet{fox2020hyperbolic}, and demonstrated to be a key ingredient in restoring hyperbolicity of the compressible two-fluid equations (refer back to Sec.~\ref{sec:ill}). In-lieu of treating the force as a separate term in Eq.~\eqref{eq:BBO}, the added mass is transported with the particle velocity, which implicitly captures the force history without the need for a response kernel. This is based on the formulation of \citet{cook1984virtual}, but generalized to a compressible fluid and a non-constant added-mass function to capture volume fraction effects. The mass appearing in the conservation equations are augmented to include the portion of the wake moving with the particle. In an EE framework, this involves replacing $\phi_p$ in the particle-phase equations with $\phi_p^\star=\phi_p+\phi_a$, such that the particles carry an added mass $m_p=(\rho_p\phi_p+\rho_f\phi_a)V_p=\rho_e\phi_p^\star V_p$, where $\alpha_a$ is the added volume fraction of the surrounding fluid and $\rho_e$ is the effective density of the particle. 
A simple model was proposed based on a mass exchange function that acts as a source term for $\phi_p^\star$, given by
\begin{equation}
    \frac{\partial \rho_e\phi_p^\star}{\partial t}+\nabla\bcdot\left(\rho_e\phi_p^\star\bm{u}_p\right)=\frac{\rho_f\phi_f\phi_p}{\tau_a}\left(C_m^\star-C_m \right),
\end{equation}
where $\tau_a$ is a time scale that characterizes the expansion/contraction/formation of particle wakes, and $C_m^\star$ is the added mass function that should depend on $\Rey_p$, $\Mac_p$, and $\phi_p$. When a particle moves from a region of high $\phi_p$ to low $\phi_p$ (i.e., into a region with larger inter-particle spacing), $C_m<C_m^\star$ and the wake will grow by entraining the surrounding fluid. In a EL framework, this would require replacing $m_p$ in Eq.~\eqref{eq:BBO} with an effective mass $m_p^\star$ carried by each particle that evolves according to
\begin{equation}
    \frac{{\rm d}m_p^\star}{{\rm d}t} = \frac{1}{\tau_a}\rho_f \phi_f V_p(C_m^\star-C_m).
\end{equation}
Such a treatment for the added mass has advantages over the more traditional approach given by Eq.~\eqref{eq:Fiu}, as the instantaneous acceleration of each phase need not be evaluated at the particle location, and the time-history of the response kernel is not required. Models for $\tau_a$ and $C_m^\star$ are needed for relevant values of $\Rey_p$, $\Mac_p$, and $\phi_p$, which could be evaluated from PR--DNS \citep[e.g., see][for a recent evaluation of $\tau_a$]{moore2019lagrangian}.

\section{Pseudo-turbulence in multi-particle systems}\label{sec:PTKE}
Flow through a collection of particles gives rise to wakes that interact non-linearly with each other and with surrounding particles. The collective effect of these flow disturbances acts as a source of fluid-phase turbulence. Because such fluctuations exist even in laminar flow (e.g., via steady wakes), and originate at small scales, it is often termed \textit{pseudo}-turbulence. This is in contrast to classical turbulence that involves a transfer of energy from large scales of motion to small scales. Models for velocity fluctuations originating from pre-existing background turbulence (i.e., from the classical turbulence cascade) are not considered here as they have already received significant attention for single-phase flow \citep[e.g.,][]{garnier2009large}. Instead, we focus on models for the pseudo-turbulent Reynolds stress, $\bm{R}_f$ in Eqs.~\eqref{eq:momentum}--\eqref{eq:energy},  that is expected to play a role when the inter-particle spacing is comparable to the size of the wakes (i.e., at moderate values of $\phi_p$).

\begin{figure}[h]
\centering
\includegraphics[width=0.9\textwidth]{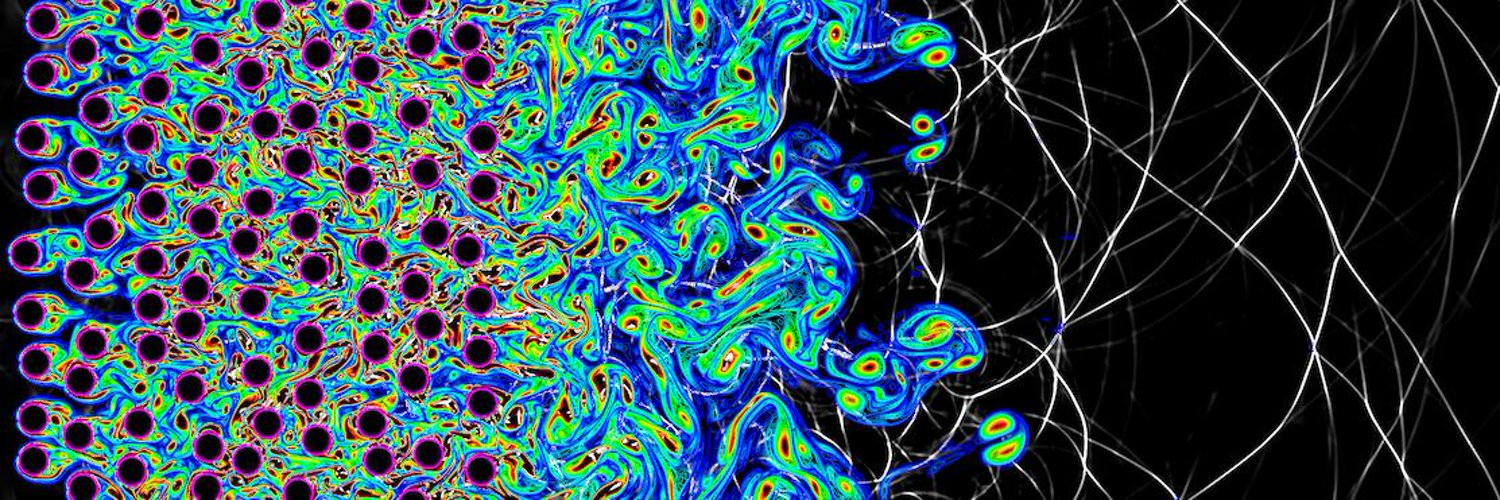}
\caption{A two-dimensional simulation showing gas-phase vorticity (color) and dilatation (black/white) shortly after a planar shock wave passes over a suspension of particles with $\Mac_s=1.22$ and $\phi_p=0.17$. Adapted from \citet{shallcross2020volume}.}
\label{fig:PTKE}
\end{figure}

When a shock wave passes over a cloud of particles at moderate volume fractions, small-scale velocity fluctuations are produced in particle interstitial sites and advected downstream with the mean flow (see Fig.~\ref{fig:PTKE}). 
\citet{theofanous2018shock} performed one of the first three-dimensional particle-resolved simulations of shock-induced dispersion. The gas phase was shown to choke near the downstream edge of the particle cloud due to the abrupt change in volume fraction, resulting in supersonic expansion that significantly increases particle acceleration. \citet{shallcross2020volume} later showed that including $\bm{R}_f$ in the definition of total fluid energy is critical for capturing this choking behavior in simulations based on averaged equations. Near the edge of the particle cloud, large gradients in $\bm{R}_f$ can be on the same order as the forces acting on the particles \citep{osnes2019computational}. The corresponding root-mean-square velocity fluctuations have been observed to be as much as 60\% of the mean flow. Its strength increases with increasing values of $\phi_p$ and incident shock Mach number \citep{mehta2019Pseudo,osnes2019computational}.

Modeling the pseudo-turbulent Reynolds stress has only gained considerable attention in recent years. An algebraic model for the pseudo-turbulent kinetic energy (PTKE), $k_f={\rm tr}(\bm{R}_f)/2$, was first proposed by \citet{mehrabadi2015pseudo} for incompressible homogeneous gas--solid flows, given by
\begin{equation}\label{eq:k1}
    \frac{k_f}{K_f}=2\phi_p+2.5\phi_p\phi_f^3\exp\left(-\phi_p\Rey_p^{1/2}\right),
\end{equation}
where $K_f=\bm{u}_f\bcdot\bm{u}_f/2$ is the kinetic energy of the resolved flow field. The correlation was developed in the range $0.1\le\phi_p\le0.5$ and $0.01\le\Rey_p\le300$ and tends to appropriate values in the zero volume fraction limit (i.e., $k_f\rightarrow 0$ when $\phi_p\rightarrow 0$). The pseudo-turbulent Reynolds stress tensor is reconstructed from the PTKE according to
\begin{equation}\label{eq:Rij}
\bm{R}_f' = 2 \rho_f k_f \left( \bm{b} +\frac{1}{3} \mathbb{I} \right),
\end{equation}
where $\bm{R}_f'$ is the Reynolds stress aligned with the local velocity difference between the phases and $\bm{b}$ is the anisotropic stress tensor that depends on $\Rey_p$ and $\phi_p$~\citep{mehrabadi2015pseudo}. $\bm{R}_f$ is then obtained by rotating $\bm{R}_f'$ to align with the Cartesian coordinate system. Details on the implementation of the rotation matrix in an EE framework can be found in \citet{peng2019implementation}. 

Using PR--DNS of shock-induced flow through random arrays of particles, \citet{osnes2019computational} proposed a model for the streamwise component of the Reynolds stress (denoted here as direction $1$) according to
\begin{equation}
    R_{f,11}=u_{f,1}^2\frac{\phi_{\rm sep}}{\phi_f-\phi_{\rm sep}},
\end{equation}
where $\phi_{\rm sep}$ represents the volume fraction of separated flow in particle wakes, which increases significantly as $\phi_p$ is reduced as a result of increasing inter-particle separation. The authors proposed a simple model of the form $\phi_{\rm sep}\left(\phi_f,\Rey_p\right)=\phi_p C\left(\Rey_p\right)$, where $C\left(\Rey_p\right)\approx 1.5$ was determined for incident shock Mach numbers between $2.2\le \Mac_s\le 3$ and volume fractions $0.05\le\phi_p\le0.1$. 
The authors note that it would be appropriate to introduce a time-dependency for $\phi_{\rm sep}$, since particle wakes and fluctuations are not generated instantaneously after the shock wave passes over a particle. This is conceptually similar to the added mass formulation proposed by \citet{fox2020hyperbolic} reported in Sec.~\ref{sec:Fiu}. However, such models do not yet exist.

A key drawback with the algebraic models proposed by \citet{mehrabadi2015pseudo} and \citet{osnes2019computational} is that they only predict finite PTKE in the regions where particles are present ($\phi_p>0$). While these fluctuations originate in the vicinity of particles, they can be transported with the mean flow into regions devoid of particles where $\phi_p=0$ (see Fig.~\ref{fig:PTKE}). Another challenge is distinguishing the sub-grid scale velocity fluctuations originating from pre-existing turbulence (via a classical energy cascade) and those induced by particles (pseudo turbulence). With this in mind, \citet{shallcross2020volume} rigorously derived a transport equation for the PTKE based on volume filtering, given by
\begin{equation}\label{eq:k}
\frac{\partial \phi_f \rho_f k_f}{\partial t} + \nabla \bcdot \left(\phi_f \rho_f \bm{u}_f k_f \right) = -\phi_f \bm{R}_f \bcol\nabla \bm{u}_f + \left(\bm{u}_p - \bm{u}_f \right) \bcdot \bm{F}_p   - \phi_f \rho_f \varepsilon_{PT},
\end{equation}
where the viscous and sub-filtered contributions are absorbed into $\varepsilon_{PT}$, which represents dissipation of PTKE. The first term on the right-hand side of Eq.~\eqref{eq:k} represents production due to mean shear--the usual turbulence production term in single phase flow. The second term on the right-hand side represents production due to particle drag that is only active when $\phi_p>0$. 
It was found that reconstructing $\bm{R}_f$ from the transported PTKE via Eq.~\eqref{eq:Rij} yields accurate predictions of the anisotropy for shock-particle interactions, despite the coefficients being derived for steady, incompressible flows. However, the results were shown to depend strongly on $\varepsilon_{PT}$. 

Following what is typically done in single-phase turbulence modeling \citep{vassilicos2015dissipation}, the dissipation rate can be modeled as $\varepsilon_{PT}\propto k_f/\tau_\varepsilon$, where $\tau_\varepsilon$ is a dissipation time scale that requires modeling. \citet{vartdal2018using} proposed a simple model of the form $\tau_\varepsilon=l/\sqrt{k_f}$, where the length scale $l$ can be related to $d_p$ or the inter-particle spacing. \citet{shallcross2020volume} proposed a model that blends two time scales: one similar to the model by \citet{vartdal2018using} away from particles; and another based on the relative velocity between the phases in the vicinity of particles. However, the model has only been tested under limited flow conditions. For broader applicability, a transport equation for the dissipation rate could be derived as well. Such an approach would of course lead to additional terms that require models. The increase in available compressible DNS data in recent years~\citep[e.g.,][]{theofanous2017dynamics, theofanous2018shock, mehta2019Pseudo,shallcross2020volume} represent an exciting opportunity for developing such models valid across relevant two-phase flow conditions.

\section{Conclusions}
Above we present a survey of available models for high-speed gas--particle flows, with an emphasis on fluid-particle interactions. This class of flows gives rise to rich multiphase flow physics that pose significant challenges in predicting its behavior. While such flows are present in numerous applications, this article was primarily motivated by PSI during planetary and lunar landings. With future planned sample-return missions from Mars and eventual crewed missions to the Moon, Mars, and beyond, the risks associated with PSI are paramount. Simulating the harsh environment during touchdown using physics-based (EE and EL) approaches offers many advantages over the use of empirical correlations for macroscopic behavior like crater depth and erosion rate. However, the subgrid-scale models they rely on are largely based on data from incompressible flows or outdated experimental measurements.

\begin{figure}[h]
\subfigure[]{\includegraphics[width=0.47\textwidth]{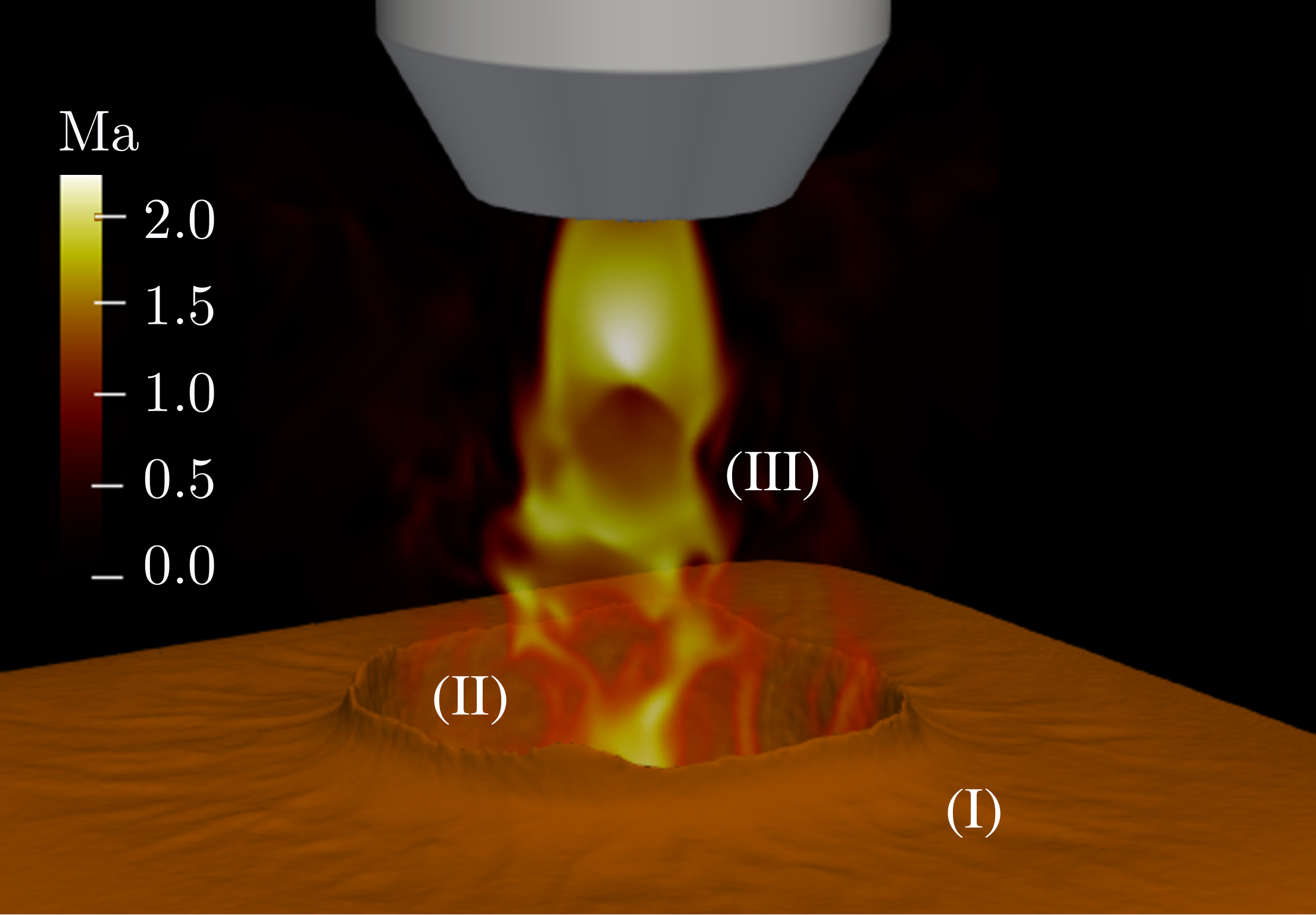}\label{fig:3DPSI}}
\subfigure[]{\includegraphics[width=0.49\textwidth]{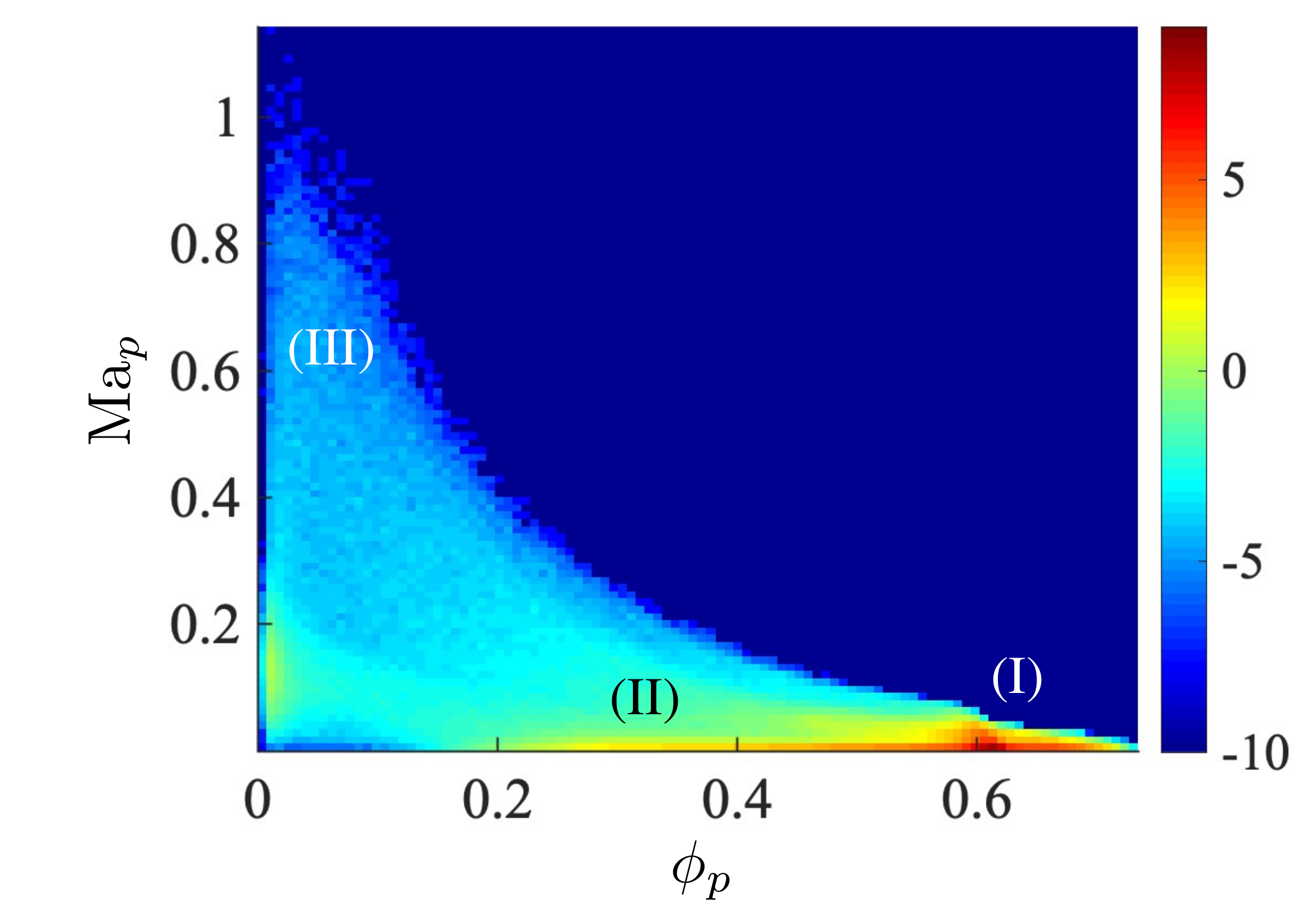}\label{fig:jPDF}}
\centering
\caption{EL simulation of an underexpanded jet impinging on a granular bed highlighting the surface (I), crater (II), and ejecta (III). (a) Visualization of the granular surface ($\phi_p=0.6$ contour shown in brown) and local gas-phase Mach number (red/yellow). (b) Joint-PDF (in log scale) of particle Mach number and volume fraction at the same instant. Adapted from \citet{shallcross2021modeling}.}
\label{fig:PSI-EL}
\end{figure}


It should be noted that when the particle phase is densely packed (i.e., near close packing $\phi_p>0.6$), a significant amount of power would be required to sustain a mean velocity difference between the phases at moderate Mach numbers. To highlight this, Fig.~\ref{fig:PSI-EL} shows results from a recent EL simulation of an underexpanded jet impinging on a bed of monodisperse spherical particles \citep{shallcross2021modeling}. The joint probability density function (PDF) of $\Mac_p$ and $\phi_p$ reveals high probability events of moderate Mach numbers at low $\phi_p$ (corresponding to the ejecta above the crater), whereas $\phi_p>0.4$ predominantly occurs when $\Mac_p<0.1$. Thus, it \textit{might} be possible to restrict model development at finite Mach numbers to dilute and moderately dense concentrations only.

Finally, it should be noted that much of the flow physics discussed throughout this review are also present in many systems on Earth. For example, pyroclastic density currents are dangerous multiphase flows emanating from volcanic eruptions that operate under a wide range of Mach numbers and volume fractions \citep{lube2020multiphase}. In addition, the detonation of a heterogeneous explosive results in rapid dispersal of high-speed solid particles \citep{zhang2001explosive}, which can have tragic consequences as seen with the 2020 Beirut explosion \citep{guglielmi2020beirut}.


\begin{tcolorbox}[arc=0mm,colback=lyellow,colframe=lyellow]
\subsection*{Summary points}
\begin{enumerate}
\item Compared to its incompressible counterpart, compressible gas--particle flows introduce new scales of motion (e.g., the shock wave thickness and acoustic time scale) that pose significant modeling challenges.
    \item The two-fluid equations for disperse multiphase systems are well known to be ill-posed. After more than four decades of attempts to remedy this, it was only recently rigorously resolved by \citet{fox2020hyperbolic}, who showed that inclusion of the added mass and a fluid-mediated contribution to the particle-phase pressure tensor are needed to ensure hyperbolicity for arbitrary density ratios.
    \item The pseudo-turbulent Reynolds stress plays an important role at moderate volume fractions, and can contribute to a significant portion of the total kinetic energy during shock-particle interactions. It is also needed to ensure conservation.
    \item Existing models for the quasi-steady drag force can be broadly categorized into (i) single-particle correlations across Reynolds and Mach numbers; and (ii) multi-particle correlations exclusively developed for incompressible flow. The intersection between these two regimes, illustrated in Fig.~\ref{fig:drag}, remain elusive. 
    \item Mach number dependent drag laws used today have surprising origins from 18th- and 19th-century cannon fire experiments.
    \item \citet{loth2021supersonic} was the first to incorporate data from numerical simulations to refine the $\Mac_p-\Rey_p$ drag coefficient for an isolated sphere where large experimental uncertainty exists. 
    \item Unlike in incompressible gas--solid flows, particle motion in a compressible flow undergoing strong acceleration (e.g., in the presence of a shock wave) can be greatly influenced by unsteady forces. 
\end{enumerate}
\end{tcolorbox}

\begin{tcolorbox}[arc=0mm,colback=lred,colframe=lred]
\subsection*{Future issues}
\begin{enumerate}
    \item With the advent of PR--DNS, the past two decades have seen an explosion of multi-particle drag correlations for incompressible flows. With PR--DNS of compressible gas--particle flows starting to come online,
    We anticipate similar progress to be made for drag models at finite $\Rey_p$, $\phi_p$, \textit{and} $\Mac_p$.
    \item Following the form given in Eq.~\eqref{eq:WenYu}, a natural choice for developing improved multi-particle compressible drag correlations would be to use the recent Reynolds and Mach number-dependent drag coefficient of \citet{loth2021supersonic} for the drag acting on an isolated particle, and augment it with the Reynolds- and volume fraction-dependent correlation of \citet{tenneti2011drag}. Its utility would need to assessed by PR--DNS.
    \item It is now recognized that flow past a collection of particles exhibits a distribution in drag forces with significant variance \citep{akiki2016force,mehta2019effect,lattanzi2020stochastic}. Various models have recently been proposed to capture higher-order drag force statistics arising from neighbor-induced flow perturbations in EL \citep{akiki2017pairwise,esteghamatian2018stochastic,seyed2020microstructure,lattanzi2021stochastic} and EE \citep{lattanzi2021fluid} methods. Future research should leverage and extend these models to the compressible flow regime.
    \item Treating the hydrodynamic force as a stochastic variable offers some potential advantages over the classical BBO treatment of modeling each force contribution separately. Rather than attempting to tease out how each pair-wise neighbor interaction contributes to the hydrodynamic force on a given particle,  \citet{lattanzi2020stochastic} demonstrated that the statistics obtained from treating the force as a stochastic variable are reconcilable with PR--DNS.
    \item While models for disperse two-phase flows have traditionally focused on hydrodynamic interactions, many applications of compressible gas--particle flows operate at elevated temperatures where thermal effects become important. Rapid changes in particle temperature can result in melting or initiate chemical reactions. Thus, improved models for inter-phase heat transfer are needed. Progress in this area can be found in \citet{ling2016inter}.
    \item This article focused exclusively on models for \textit{spherical} particles. Lunar and martian regolith is often polydisperse, and can be highly non spherical. This has important consequences on fluid-particle interactions (e.g., drag and lift) in addition to particle-particle interactions (e.g., internal friction, angle of repose). Special care needs to be taken to incorporate such effects in a manner that does not result in introducing `tuning' parameters that can be used to fit data.
    \item EE methods rely on models for particle-phase rheology obtained from kinetic theory of granular flows. Future research is warranted to validate or extend these models to compressible flows.
\end{enumerate}
\end{tcolorbox}

\section*{Acknowledgements} 
The author thanks Professor Jason Rabinovitch for fruitful discussions and comments related to PSI, and Drs. Magnus Vartdal and Andreas Osnes for providing critical feedback on various modeling aspects. This work was supported by the National Aeronautics and Space Administration (NASA) under grant number 80NSSC20K1868.

\appendix

\section{Some common drag coefficients}
\label{app:drag}
In this Appendix, various correlations for the drag coefficient are provided as \code{Matlab} functions, along with their range of validity and general comments.

\subsection{\citet{clift1971motion}}\label{app:CG}
\begin{table}[H]
\begin{tabular}{ll}
Validity & Comment\\
\midrule
$\Rey_p<2\times10^5$ & \multirow[t]{4}{12cm}{Standard Reynolds number-dependent drag coefficient used for an isolated particle in incompressible flow. Reduces to the classical Schiller--Naumann correlation at moderate Reynolds numbers and to Stokes drag when $\Rey_p\rightarrow 0$.}\\
$\Mac_p=0$ & \\
$\phi_p=0$ & \\
\end{tabular}
\end{table}
\begin{lstlisting}
function CD = Clift_Gauvin(Rep)
CD = 24 / Rep * (1 + 0.15 * Rep^(0.687)) + 0.42 / (1 + 42500 / Rep^(1.16));
end
\end{lstlisting}

\subsection{\citet{henderson1976drag}}\label{app:henderson}
\begin{table}[H]
\begin{tabular}{ll}
Validity & Comment\\
\midrule
$\Rey_p<2\times10^4$ & \multirow[t]{5}{12cm}{In addition to the usual Reynolds and Mach number based on the relative velocity, the drag correlation is also a function of the Reynolds and Mach number based on the freestream velocity, $\Rey_\infty$ and $\Mac_\infty$. It also includes effects of the temperature ratio between the gas and particle $T_p/T_f$.}\\
$\Mac_p<6$ & \\
$\phi_p=0$ & \\
&\\
\end{tabular}
\end{table}
\begin{lstlisting}
function CD = Henderson(Rep,Map,Re_inf,Ma_inf,Tf,Tp)
global gamma
S=Map*sqrt(0.5*gamma);
S_inf=MaInf*sqrt(0.5*gamma);
if Ma <= 1
    CD = 24 / (Rep + Map*sqrt(0.5*gamma)*(4.33 + (3.65 - 1.53*(Tp/Tf))/(1 + 0.353*(Tp/Tf))) * exp(-0.247*Rep/(S)))+ exp(-0.5*Map/sqrt(Rep))*((4.5 + 0.38 * (0.03*Rep + 0.48*sqrt(Rep))) / (1 + 0.03*Rep + 0.48*sqrt(Rep)) + 0.1*Map^2 + 0.2*Map^8) + (1 - exp(-1*Map/Rep)) * 0.6 * S;
elseif Map > 1 && Map < 1.75
    CD = (24 / (Rep + sqrt(0.5*gamma) * (4.33 + (3.65 - 1.53*(Tp/Tf)) / (1 + 0.353*(Tp/Tf))) * exp(-0.247*Rep / (sqrt(0.5*gamma)))) + exp(-0.5*1/sqrt(Rep)) * ((4.5 + 0.38 * (0.03*Rep + 0.48*sqrt(Rep))) / (1 + 0.03*Rep + 0.48*sqrt(Rep))+ 0.1*1^2 + 0.2*1^8) + (1 - exp(-1/Rep)) * 0.6 * sqrt(0.5*gamma)) + 4/3 * (Ma_inf - 1) * ((0.9 + 0.34/1.75^2 + 1.86 * sqrt(1.75/Re_inf) * (2 + 2/(1.75*sqrt(0.5*gamma))^2+ 1.058 / (1.75*sqrt(0.5*gamma)) * sqrt(Tp/Tf) - 1 / (1.75*sqrt(0.5*gamma))^4)) / (1 + 1.86 * sqrt(1.75/Re_inf)) - (24 / (Rep + sqrt(0.5*gamma) * (4.33 + (3.65 - 1.53*(Tp/Tf)) / (1 + 0.353*(Tp/Tf))) * exp(-0.247*Rep/(sqrt(0.5*gamma)))) + exp(-0.5*1/sqrt(Rep))*((4.5 + 0.38 * (0.03*Rep + 0.48*sqrt(Rep))) / (1 + 0.03*Rep + 0.48*sqrt(Rep))+ 0.1*1^2 + 0.2*1^8) + (1 - exp(-1/Rep)) * 0.6 * sqrt(0.5*gamma)));
else
    CD = (0.9 + 0.34/Ma_inf^2 + 1.86 * sqrt(Ma_inf/Re_inf) * (2 + 2/(S_inf)^2 + 1.058/(S_inf) * sqrt(Tp/Tf) - 1/(S_inf)^4))/ (1 + 1.86 * sqrt(Ma_inf/Re_inf));
    
end
end
\end{lstlisting}

\subsection{\citet{parmar2010improved}}
\label{app:parmar}
\begin{table}[H]
\begin{tabular}{ll}
Validity & Comment\\
\midrule
$\Rey_p\le2\times10^5$ & \multirow[t]{4}{12cm}{The original paper \citep{parmar2010improved} contains a typo, the correct formulation can be found in the subsequent thesis \citep{parmar2010unsteady}. The original correlation was developed for an isolated particle, \citet{ling2012interaction} added a correction for $0\le\phi_p\le0.3$.}\\
$\Mac_p\le1.75$ & \\
$\phi_p=0$ & \\
\end{tabular}
\end{table}
\begin{lstlisting}
function CD = Parmar(Rep,Map,phip)
fsub=zeros(1,3);
fsup=zeros(1,3);
Csup=zeros(1,3);
Csub=zeros(1,3);

Mcr = 0.6;
CDMcr = 24 / Rep * (1 + 0.15 * Rep^(0.684)) + 0.513 / (1 + 483 / Rep^(0.669));
CDM1   = 24 / Rep * (1 + 0.118 * Rep^(0.813)) +  0.69 / (1 + 3550 / Rep^(0.793));
CDM175 = 24 / Rep * (1 + 0.107 * Rep^(0.867)) + 0.646 / (1 + 861 / Rep^(0.634));

if Map <= Mcr
    % Subcritical regime (standard drag of Clift and Gauvin)
    CDstd = 24 / Rep * (1 + 0.15 * Rep^(0.687)) + 0.42 / (1 + 42500 / Rep^(1.16));
    CD = CDstd + (CDMcr - CDstd) * Map / Mcr;
elseif Map>Mcr && Map<=1
    % Intermediate regime (supercritical and subsonic)
    Csub(1) = 6.48; Csub(2) = 9.28; Csub(3) = 12.21;
    fsub(1) = -0.087 + 2.92*Map - 4.75*Map^2 + 2.83*Map^3;
    fsub(2) = -0.12  + 2.66*Map - 4.36*Map^2 + 2.53*Map^3;
    fsub(3) =  1.84 - 5.13*Map + 6.05*Map^2 - 1.91*Map^3;
    fsubM1(1) = 0.913; fsubM1(2) = 0.71; fsubM1(3) = 0.85;
    fsubMcr(1) = 0.56628; fsubMcr(2) = 0.45288; fsubMcr(3) = 0.52744;
    zsub =   ...
        (fsub(1) - fsubMcr(1)) / (fsubM1(1) - fsubMcr(1)) * (log(Rep) - Csub(2))/(Csub(1)-Csub(2)) * (log(Rep)-Csub(3))/(Csub(1)-Csub(3)) + ...
        (fsub(2) - fsubMcr(2)) / (fsubM1(2) - fsubMcr(2)) * (log(Rep) - Csub(1))/(Csub(2)-Csub(1)) * (log(Rep)-Csub(3))/(Csub(2)-Csub(3)) + ...
        (fsub(3) - fsubMcr(3)) / (fsubM1(3) - fsubMcr(3)) * (log(Rep) - Csub(1))/(Csub(3)-Csub(1)) * (log(Rep)-Csub(2))/(Csub(3)-Csub(2));
    CD = CDMcr + (CDM1 - CDMcr) * zsub;
elseif Map > 1 && Map<1.75
    % Supersonic regime
    Csup(1) = 6.48; Csup(2) = 8.93; Csup(3) = 12.21;
    fsup(1) = 0.126 + 1.15*Map - 0.306*Map^2 - 0.007*Map^3 - 0.061*exp((1-Map)/0.011);
    fsup(2) = -0.901 + 2.93*Map - 1.573*Map^2 + 0.286*Map^3 - 0.042*exp((1-Map)/0.01);
    fsup(3) = 0.13 + 1.42*Map - 0.818*Map^2 + 0.161*Map^3 - 0.043*exp((1-Map)/0.012);
    fsupM1(1) = 0.902; fsupM1(2) = 0.7; fsupM1(3) = 0.85;
    fsupM175(1) = 1.163859375; fsupM175(2) = 0.94196875; fsupM175(3) = 0.972734375;
    zsup = ...
        (fsup(1) - fsupM1(1)) / (fsupM175(1) - fsupM1(1)) * ...
        (log(Rep) - Csup(2))/(Csup(1)-Csup(2)) * (log(Rep)-Csup(3))/(Csup(1)-Csup(3)) + ...
        (fsup(2) - fsupM1(2)) / (fsupM175(2) - fsupM1(2)) * ...
        (log(Rep) - Csup(1))/(Csup(2)-Csup(1)) * (log(Rep)-Csup(3))/(Csup(2)-Csup(3)) + ...
        (fsup(3) - fsupM1(3)) / (fsupM175(3) - fsupM1(3)) * ...
        (log(Rep) - Csup(1))/(Csup(3)-Csup(1)) * (log(Rep)-Csup(2))/(Csup(3)-Csup(2));
    CD = CDM1 + (CDM175 - CDM1) * zsup;
    else
    % Map > 1.75 is outside validity
    disp('Map>1.75 not valid')
end

% Correct for volume fraction (Sangani et al., 1991)
CD = CD * (1 + 2 * phip) / (1 - phip)^2;
end
\end{lstlisting}

\subsection{\citet{tenneti2011drag}}\label{app:tenneti}
\begin{table}[H]
\begin{tabular}{ll}
Validity & Comment\\
\midrule
$0.01\le\Rey_p<300$ & \multirow[t]{8}{12cm}{Developed using PR--DNS of homogeneous flow past random assemblies of fixed spherical particles. Refinement using freely-evolving particles with $0.001\le\rho_p/\rho_f\le1000$ can be found in \citet{tavanashad2021particle}. The correlation for freely-evolving particles approaches that of fixed particle assemblies when $\rho_p/\rho_f>100$. Note there is a difference in a factor $(1-\phi_p)$ from the original paper \citep{tenneti2011drag} in order to remove the average pressure gradient used to force the flow.}\\
$\Mac_p=0$ & \\
$0.1\le\phi_p\le0.5$ & \\
&\\
&\\
&\\
&\\
\end{tabular}
\end{table}
\begin{lstlisting}
function CD = Tenneti(Rep,phip)
b1 = 5.81*phip/(1-phip)^3+0.48*phip^(1/3)/(1-phip)^4;
b2 = phip^3*Rep*(0.95+0.61*phip^3/(1-phip)^2);
F = (1-phip) * ((1+0.15*Rep^(0.687))/(1-phip)^3+b1+b2);
CD=24*F/Rep;
end
\end{lstlisting}

\subsection{\citet{tang2016direct}}\label{app:tang}
\begin{table}[H]
\begin{tabular}{ll}
Validity & Comment\\
\midrule
$40\le\Rey_p<1000$ & \multirow[t]{8}{12cm}{Developed using PR--DNS of flow past freely-evolving spherical particles with elastic collisions. In addition to the usual Reynolds number, the drag correlation depends on a Reynolds number based on the granular temperature $\Theta_p$, $\Rey_T=\rho_f\sqrt{\Theta_p}d_p/\mu_f$, to account for the effect of particle mobility. The contribution of particle mobility is accounted for though the force deviation $\Delta F_D=2.98\Rey_T\phi_p/(1-\phi_p)^2$. This is combined with the drag for static array of particles proposed by \citet{tang2015new}.}\\
$\Mac_p=0$ & \\
$0.1\le\phi_p\le0.45$ & \\
$\rho_p/\rho_f>500$ & \\
&\\
&\\
&\\
\end{tabular}
\end{table}
\begin{lstlisting}
function CD = Tang(Rep,phip,Re_T)
F = 10*phip/(1-phip)^2 + (1-phip)^2 * (1+1.5*sqrt(phip)) + (0.11*phip*(1+phip) - 0.00456/(1-phip)^4 + (0.169*(1-phip) + 0.0644/(1-phip)^4) * Rep^(-0.343)) * Rep + 2.98*Re_T * phip/(1-phip)^2;
CD=24*F/Rep;
end
\end{lstlisting}

\subsection{\citet{loth2021supersonic}}\label{app:loth}
\begin{table}[H]
\begin{tabular}{ll}
Validity & Comment\\
\midrule
$\Rey_p\le2\times10^5$ & \multirow[t]{4}{12cm}{A refinement of the popular \citet{loth2008compressibility} drag correlation using an expanded experimental dataset along with DNS of (continuum) supersonic flow past a sphere and rarefied gas simulations via direct simulation Monte Carlo (DSMC).}\\
$\Mac_p<10$ & \\
$\phi_p=0$ & \\
\end{tabular}
\end{table}
\begin{lstlisting}
function CD = Loth(Rep,Map)
global gamma
eps = 1e-12;
if Rep <= 45
    % Rarefraction-dominated regime
    Knp = sqrt(0.5 * pi * gamma) * Map / Rep;
    fKn = 1 / (1 + Knp*(2.514 + 0.8*exp(-0.55/Knp)));
    CD1 = 24 / Rep * (1 + 0.15*Rep^(0.687)) * fKn;
    s = Map * sqrt(0.5 * gamma);
    if (Map <= 1)
        JM = 2.26 - 0.1/Map + 0.14/Map^3;
    else
        JM = 1.6 + 0.25/Map + 0.11/Map^2 + 0.44/Map^3;
    end
    CD2 = (1 + 2*s^2) * exp(-s^2) / (s^3*sqrt(pi) + eps) + (4*s^4 + 4*s^2 - 1) * erf(s) / (2*s^4 + eps) + 2 / (3 * s + eps) * sqrt(pi);
    CD2 = CD2 / (1 + (CD2/JM - 1) * sqrt(Rep/45));
    CD = CD1 / (1 + Map^4) + Map^4 * CD2 / (1 + Map^4);
else
    % Compression-dominated regime
    if (Map < 1.5)
        CM = 1.65 + 0.65 * tanh(4*Map - 3.4);
    else
        CM = 2.18 - 0.13 * tanh(0.9*Map - 2.7);
    end
    if (Map < 0.8)
        GM = 166*Map^3 + 3.29*Map^2 - 10.9*Map + 20;
    else
        GM = 5 + 40*Map^(-3);
    end
    if (Map < 1)
        HM = 0.0239*Map^3 + 0.212*Map^2 - 0.074*Map + 1;
    else
        HM = 0.93 + 1 / (3.5 + Map^5);
    end
    CD = 24/Rep*(1 + 0.15 * Rep^(0.687))*HM + 0.42*CM / (1+42500/Rep^(1.16*CM) + GM/sqrt(Rep));
end
end
\end{lstlisting}

\bibliographystyle{model2-names}
\bibliography{main}

\end{document}